\documentclass[superscriptaddress,10pt,onecolumn, accepted=2019-07-08]{quantumarticle}
\pdfoutput=1
\usepackage[numbers, sort&compress, merge]{natbib}

\usepackage[utf8]{inputenc}
\usepackage[english]{babel}
\usepackage[T1]{fontenc}
\usepackage{amsmath}
\usepackage[bookmarks=false]{hyperref}

\usepackage{complexity}

\usepackage{amsthm}
\usepackage{amsmath}
\usepackage{amsfonts,amssymb}
\usepackage{graphicx}
\usepackage{longtable}
\usepackage{multirow} % multiple rows in table
\usepackage{algorithmicx}
\usepackage{algorithm}
\usepackage{afterpage}
\usepackage{enumitem}
\usepackage{comment}
\usepackage{epstopdf} % For images
\usepackage{bbold} %For Identity operator
\usepackage{tabularx}
\newcolumntype{D}{>{\centering\arraybackslash}X}
\newcommand{\myvspace}{\rule{0pt}{3ex}}

\usepackage{thmtools}
\usepackage{amsthm}
\usepackage{amsmath}
\usepackage{amsfonts,amssymb}
\usepackage{calc}
\usepackage{tikz}
\usepackage{epstopdf} % For images
\usepackage{setspace}

\newcommand{\ket}[1]{|#1\rangle}
\newcommand{\bra}[1]{\langle#1|}
\newcommand{\stab}[1]{\langle#1\rangle}
\newcommand{\I}{\hat{\mathcal{I}}}
\newcommand{\tr}[1]{\operatorname{Tr}\left[#1\right]}

\newcommand{\s}{\text{span}}

\newtheorem{theorem}{Theorem}
\newtheorem{definition}{Definition}
\newtheorem{lemma}[theorem]{Lemma}
\newtheorem{corollary}[theorem]{Corollary}

\usepackage[capitalise,nameinlink]{cleveref}

\def\undertilde#1{\mathord{\vtop{\ialign{##\crcr
				$\hfil\displaystyle{#1}\hfil$\crcr\noalign{\kern1.5pt\nointerlineskip}
				$\hfil\tilde{}\hfil$\crcr\noalign{\kern1.5pt}}}}}

\begin{document}
\title{Hamiltonian Simulation by Qubitization}
\date{\today}
\author{Guang Hao Low}
\orcid{0000-0002-6934-1052}
\affiliation{Department of Physics, Massachusetts Institute of Technology, Cambridge, Massachusetts, USA}
\author{Isaac L. Chuang}
\affiliation{Department of Electrical Engineering and Computer Science, Department of Physics, Research Laboratory of Electronics, Massachusetts Institute of Technology, Cambridge, Massachusetts, USA}

\maketitle
\begin{abstract}
We present the problem of approximating the time-evolution operator $e^{-i\hat{H}t}$ to error $\epsilon$, where the Hamiltonian $\hat{H}=(\langle G|\otimes\I)\hat{U}(|G\rangle\otimes\I)$ is the projection of a unitary oracle $\hat{U}$ onto the state $|G\rangle$ created by another unitary oracle. Our algorithm solves this with a query complexity $\mathcal{O}\big(t+\log({1/\epsilon})\big)$ to both oracles that is optimal with respect to all parameters in both the asymptotic and non-asymptotic regime, and also with low overhead, using at most two additional ancilla qubits. This approach to Hamiltonian simulation subsumes important prior art considering Hamiltonians which are $d$-sparse or a linear combination of unitaries, leading to significant improvements in space and gate complexity, such as a quadratic speed-up for precision simulations. It also motivates useful new instances, such as where $\hat{H}$ is a density matrix. A key technical result is `qubitization', which uses the controlled version of these oracles to embed any $\hat{H}$ in an invariant $\text{SU}(2)$ subspace. A large class of operator functions of $\hat{H}$ can then be computed with optimal query complexity, of which $e^{-i\hat{H}t}$ is a special case.
\end{abstract}
\vspace{-1.0cm}
\tableofcontents
\newpage
\section{Introduction}
Quantum computers were originally envisioned as machines for efficiently simulating quantum Hamiltonian dynamics. As Hamiltonian simulation is \BQP-complete, the problem is believed to be intractable by classical computers, and remains a strong primary motivation.
The first explicit quantum algorithms for Hamiltonian simulation were discovered by Lloyd~\cite{Lloyd1996universal} for local interactions, and then generalized by Aharonov and Ta-Shma~\cite{Aharonov2003Adiabatic} to sparse Hamiltonians. Celebrated achievements over the years~\cite{Childs2012, Berry2012, LloydMohseniRebentrost2014, Berry2015Hamiltonian, Low2016HamSim} have each ignited a flurry of activity in diverse applications from quantum algorithms~\cite{Harrow2009,Childs2015LinearSystems,Chowdhury2016quantum,
	brandao2016quantum} to quantum chemistry~\cite{Yung2014Physics, Wecker2014, poulin2014trotter, Reiher2016Reaction, Babbush2016Exponential, Kivlichan2016bounding,Omalley2016Scalable}. In this dawning era of the small quantum computer~\cite{BarendsKellyMegrantEtAl2014,Debnath2016Demonstration}, 
%where reports of quantum gate error rates below the threshold for fault-tolerant error correction are increasingly commonplace, 
the relevance and necessity of space and gate efficient procedures for practical Hamiltonian simulation has intensified.

The cost of simulating the time-evolution operator $e^{-i\hat{H}t}$ depends on several factors: the number of system qubits $n$, evolution time $t$, target error $\epsilon$, and how information on the Hamiltonian $\hat{H}$ is accessed by the quantum computer. This field has progressed rapidly following groundbreaking work in the fractional query model~\cite{Berry2014Exponential}, which was the first to achieve query complexities that depend logarithmically on error. This was generalized by Berry, Childs, Cleve, Kothari, and Somma (BCCKS)~\cite{Berry2015Truncated} to the common case where $\hat{H}=\sum_{j=1}^d\alpha_j\hat{U}_j$ is a linear combination of $d$ unitaries\footnote{A particularly interesting and general class of systems are $k$-local Hamiltonians with $m$ terms. These are also $m2^k$-sparse, and can be expressed as a linear combination of $\le m 4^k$ Pauli operators, each of which are unitary.} 
with an algorithm using $\mathcal{O}\big(\frac{\log{(d)}\log{(\tau/\epsilon)}}{\log\log{(\tau/\epsilon)}}\big)$ ancilla qubits and only $\mathcal{O}\big(\frac{\tau\log{(\tau/\epsilon)}}{\log\log{(\tau/\epsilon)}}\big)$ queries, where $\tau\leftarrow\|\vec{\alpha}\|_1 t$ and $\|\vec{\alpha}\|_1=\sum_{j=1}^d|\alpha_j|$ is the usual $L_1$ norm. Subsequently, an extension was made to $d$-sparse Hamiltonians~\cite{Berry2015Hamiltonian} that have at most $d$ non-zero elements per row, also using $\mathcal{O}\big(\frac{\tau\log{(\tau/\epsilon)}}{\log\log{(\tau/\epsilon)}}\big)$ queries with $\tau\leftarrow dt\|H\|_{\text{max}}$. A prominent open question in all these works was whether the \emph{additive} lower bound $\Omega\big(\tau+\frac{\log({1/\epsilon})}{\log\log{(1/\epsilon)}}\big)$\footnote{A slightly tighter lower bound is $\Omega(q)$ queries, where $q=\min{\{q\in\mathbb{Z}^{+}:\epsilon<|\sin(|\tau|/q)|^q=\mathcal{O}((\tau/q)^q)\}}$~\cite{Berry2014Exponential,Berry2015Hamiltonian}. Note that $q\in\mathcal{O}(\tau+\log{(1/\epsilon)})\cap\Omega\big(\tau+\frac{\log({1/\epsilon})}{\log\log{(1/\epsilon)}}\big)$.}
was achievable for any of these models.

Recently, we found for the $d$-sparse model~\cite{Low2016HamSim}, which is popular in algorithm design by quantum walks~\cite{childs2003Exponential}, a procedure achieving the optimal trade-off between all parameters up to constant factors, with query complexity $\mathcal{O}\big(\tau+\log{(1/\epsilon)}\big)$\footnote{In the respective limits $\tau=\Theta(1)$ and $\epsilon=\Theta(1)$, the query complexity is $\mathcal{O}\big(\frac{\log({1/\epsilon})}{\log\log{(1/\epsilon)}}\big)$ and $\mathcal{O}(\tau)$. Importantly, this does not imply a complexity of $\mathcal{O}\big(\tau+\frac{\log({1/\epsilon})}{\log\log{(1/\epsilon)}}\big)$. The precise form of the upper bound matches the lower bound with respect to all parameters up to constant factors and lies between $\Omega\big(\tau+\frac{\log({1/\epsilon})}{\log\log{(1/\epsilon)}}\big)$ and $\mathcal{O}\big(\tau+\log{(1/\epsilon)}\big)$}. This strictly linear-time performance with additive complexity is a quadratic improvement over prior art for precision simulations when $t=\Theta(\log{(1/\epsilon)})$. Moreover, the number of ancilla qubits required is independent of $\tau$ and $\epsilon$, which is another important practical improvement. The approach, based on Childs'~\cite{Childs2010,Berry2012,Kothari2014efficient} extension of Szegedy's quantum walk~\cite{Szegedy2004spectra}, required two quantum oracles: one accepting a row $j$ and column $k$ index to return the value of entry $\hat{H}_{jk}$ to $m$ bits of precision, and another accepting a $j$ row and $l$ sparsity index to return in-place the $l^{\text{th}}$ non-zero entry in row $j$.

Unfortunately, the $d$-sparse model is less appealing in practical implementations for several reasons. First, it is exponentially slower than BCCKS when the $\hat{U}_j$ are of high weight with sparsity $\mathcal{O}(2^{n})$.
Second, its black-box oracles can be challenging to realize. Avoiding the exponential blowup by exploiting sparsity requires that positions of non-zero elements are both few in number and efficiently row-computable, which is not always the case. 
Third, the Childs quantum walk requires a doubling of the $n$ system qubits, which is not required by BCCKS. It is unclear whether our methodology could be applied to other formulations of Hamiltonian simulation, in contrast to alternatives that seem more flexible~\cite{Berry2016corrected}.

\begin{table}[t]
	\begin{tabular}{c|c|c|c|c|c}
		\multirow{ 2}{*}{Algorithm} & \multirow{ 2}{*}{Model} & \multirow{ 2}{*}{Ancilla qubits} & Query Complexity   & \multicolumn{2}{c}{Gates per Query $\mathcal{O}(\cdot)$} \\
		
		&&&$\mathcal{O}(\cdot)$&Oracle&Primitive\\
		\hline\hline
		LC~\cite{Low2016HamSim} & Sparse & $n+\mathcal{O}(m)$ & $d t+\frac{\log{(1/\epsilon)}}{\log\log{(1/\epsilon)}}$ & Varies & $n+\mathrm{poly}(m)$ \\
		BCCKS~\cite{Berry2015Truncated} & $\sum_j\alpha_j\hat{U}_j$ & $\mathcal{O}(\frac{\log{(d)}\log{(\|\vec{\alpha}\|_1 t/\epsilon)}}{\log\log{(\|\vec{\alpha}\|_1 t/\epsilon)}})$ & $\frac{\|\vec{\alpha}\|_1 t\log{(\|\vec{\alpha}\|_1 t/\epsilon)}}{\log\log{(\|\vec{\alpha}\|_1 t/\epsilon)}}$  & $dC$ & $\log{(d)}$\\
		LMR~\cite{LloydMohseniRebentrost2014} & Mixed $\rho$ & $n+1$ & $t^2/\epsilon$ & Varies & $\log{(n)}$\\ 
		\hline
		~\cref{thm:QueryComplexity}& $\bra{G}\hat{U}\ket{G}$ &  $\lceil\log_{2}{(d)}\rceil+2$ & $t+\frac{\log{(1/\epsilon)}}{\log\log{(1/\epsilon)}}$ & Varies & $\log{(d)}$\\
		~\cref{cor:HamSum} & $\sum_j\alpha_j\hat{U}_j$ & $\lceil\log_{2}{(d)}\rceil+2$ & $\|\vec{\alpha}\|_1 t+\frac{\log{(1/\epsilon)}}{\log\log{(1/\epsilon)}}$ & $dC$ & $\log{(d)}$\\
		~\cref{cor:HamPure} & Purified $\rho$ & $n+\lceil\log_{2}{(d)}\rceil + 2$ & $t+\frac{\log{(1/\epsilon)}}{\log\log{(1/\epsilon)}}$ & Varies & $\log{(n)}$
	\end{tabular}
	\caption{\label{Tab:Comparison} Comparison of state-of-art with our new approaches (bottom three lines) for approximating  $e^{-i\hat{H}t}$ of $\hat{H}\in\mathbb{C}^{2^n\times 2^n}$ with error $\epsilon$. The $d$-sparse simulation oracle describes entries of $\hat{H}$ with maximum absolute value $\|\hat{H}\|_{\text{max}}=1$ to $m$ bits of precision. The BCCKS oracle provides the decomposition $\hat{H}=\sum_{j=1}^d\alpha_j\hat{U}_j$, and each $\hat{U}_j$ is given a cost $\mathcal{O}(C)$. The LMR query complexity refers to samples of the density matrix $\hat{\rho}=\hat{H}$. This work generalizes the above with oracles $\hat{G}\ket{0}=\ket{G}\in\mathbb{C}^d,\hat{U}\in\mathbb{C}^{2^nd\times 2^nd}$ such that $\bra{G}\hat{U}\ket{G}=\hat{H}$, where $\|\hat{H}\|\le 1$. A new model where the oracle that outputs the purification $\ket{\rho}=\sum^d_{j=1}\alpha_j \ket{j}_a\ket{\psi_j}$, $\text{Tr}_a[\ket{\rho}\bra{\rho}]=\hat{\rho}$ is provided. }
\end{table}

Ideally, the best features of these two algorithms could be combined, such as in~\cref{Tab:Comparison}. For example, given the decomposition $\hat{H}=\sum_{j=1}^d\alpha_j\hat{U}_j$, one would like the optimal additive complexity of sparse Hamiltonian simulation, but with the BCCKS oracles that are more straightforward to implement. 
Furthermore, one could wish for a constant ancilla overhead, of say $\lceil\log_2(d)\rceil+2$, superior to either algorithm. These improvements would greatly enhance the potential of early practical applications of quantum computation.

We achieve precisely this optimistic fusion via an extremely general procedure, made possible by what we call `qubitization', that subsumes both and motivates new formulations of Hamiltonian simulation, as captured by this main theorem:
\begin{theorem}[Optimal Hamiltonian simulation by Qubitization]
	\label{thm:QueryComplexity}
	Let $(\bra{G}_a\otimes \I_s)\hat{U}(\ket{G}_a\otimes\I_s)=\hat{H}\in\mathbb{C}^{N\times N}$ be Hermitian for some unitary $\hat{U}\in\mathbb{C}^{Nd\times Nd}$ and some state preparation unitary $\hat{G}\ket{0}_a=\ket{G}_a\in\mathbb{C}^{d}$. Then $e^{-i\hat{H}t}$ can be simulated for time $t$, error $\epsilon$ in spectral norm, and failure probability $\mathcal{O}(\epsilon)$, using at most $\lceil\log_{2}(N)\rceil+\lceil\log_{2}(d)\rceil+2$ qubits in total, $\Theta(Q)$ queries to controlled-$\hat{G}$, controlled-$\hat{U}$, and their inverses, and $\mathcal{O}(Q \log{(d)})$ additional two-qubit quantum gates where
	\begin{align}
	Q=\min{\left\{q\in\mathbb{Z}^{+}:\epsilon\ge\frac{4|t|^q}{q!2^q}=\mathcal{O}\left(\Big(\frac{e\tau}{2q}\Big)^q\right)\right\}}=\mathcal{O}(t+\log{(1/\epsilon)}).
	\end{align}
\end{theorem}
The optimality of the procedure follows by using oracles $\hat{G}$ and $\hat{U}$ that implement the Childs quantum walk. Furthermore, the transparent Hamiltonian input model of~\cref{thm:QueryComplexity} significantly expedites the development of new useful formulations of Hamiltonian simulation. For instance, we easily obtain a new result for the scenario where $\hat{H}$ is a density matrix $\hat{\rho}$. Whereas $\hat{\rho}$ can be produced by discarding the ancilla of some output from a quantum circuit $\hat{G}$, we instead keep this ancilla, leading to an unconditional quadratic improvement in time scaling, and an exponential improvement in error scaling over the sample-based Lloyd, Mohseni, and Rebentrost (LMR) model~\cite{LloydMohseniRebentrost2014,kimmel2016hamiltonian}, as summarized in~\cref{Tab:Comparison}. Though certainly a stronger model than was previously considered, the inputs to many quantum machine learning as well as quantum semidefinite programming~\cite{brandao2016quantum} applications are actually of this more restricted form, and are thus enhanced.

In fact, Hamiltonian simulation is just one application of our main innovation: an approach we call the `quantum signal processor', where the equation $(\bra{G}_a\otimes \I_s)\hat{U}(\ket{G}_a\otimes \I_s)=\hat{H}$ in~\cref{thm:QueryComplexity} is interpreted as a non-unitary \emph{signal} operator $\hat{H}$ encoded in a subspace of an oracle $\hat{U}$ flagged by $\ket{G}_a\otimes\I_s$. This general encoding defined in~\cref{def:standardform}\footnote{After the original preprint release of this manuscript, standard-form encoding has been replaced in the community by `block-encoding'~\cite{Chakraborty2018BlockEncoding}, which is a more informative description.} unveils a systematic approach toward exponential quantum speedups with generic inputs, and is central to our results:
\begin{definition}[Standard-form]
	\label{def:standardform}
	A signal operator $\hat{H}$ with spectral norm $\|\hat{H}\|\le1$ is encoded in the standard-form if we may query a unitary oracle $\hat{U}:\mathcal{H}_a\otimes\mathcal{H}_s\rightarrow \mathcal{H}_a\otimes\mathcal{H}_s$ and a unitary state preparation oracle $\hat{G}\ket{0}_a=\ket{G}_a\in\mathcal{H}_a$ with the property $(\bra{G}_a\otimes \I_s)\hat{U}(\ket{G}_a\otimes \I_s)=\hat{H}$. We also assume query access to the inverses and controlled versions of $\hat{U},\hat{G}$.
\end{definition}

The inputs to many problems, highlighted in~\cref{Tab:Applications}, can be of this form. We introduce `qubitization': the essential first step that converts this description of $\hat{H}$ into unitary evolution with properties that depend directly on $\hat{H}$. When $\hat{H}$ is Hermitian, The oracles $\hat{G}$ and $\hat{U}$ are queried to obtain a Grover-like search parallelized over all eigenvalues $\lambda$ of $\hat{H}$ through the \emph{iterate}, which is isomorphic to $e^{-i\hat{Y}\otimes \cos^{-1}{[\hat{H}]}}$ in some subspace.
Similar structures are of foundational importance to many quantum algorithms -- the gap $\Delta$ of eigenvalues $\lambda = 1-\Delta$ of $\hat{H}$ is amplified to $\cos^{-1}{(\lambda )}=\mathcal{O}(\sqrt{\Delta})$ in the phase of the iterate, which resembles spectral gap amplification~\cite{Somma2013SpectralGap}, the quantization of stochastic matrices~\cite{Szegedy2004Markov}, as well as Szegedy's~\cite{Szegedy2004spectra} and Childs'~\cite{Childs2010} quantum walk. The key difference lies in the generalized encoding of the signal operator through any $\hat{G}$ and $\hat{U}$ of~\cref{def:standardform}, instead of the more restrictive sparse Hamiltonian formulation.
%, instead of via oracles of, say, the $d$-sparse formulation. 
%In short, qubitization enables the amplification of any Hamiltonian encoded in the general manner of~\cref{thm:QueryComplexity}.

Unlike Grover search, we do not seek to prepare some target state. Rather, we exploit a direct sum of Grover-like rotations that are each isomorphic to $\text{SU}(2)$, to engineer arbitrary target functions $f(\lambda)$ of its overlap $\lambda$. The quantum signal processor exploits this structure to attack the often-considered problem of designing a quantum circuit $\hat{Q}$ that queries $\hat{G}$ and $\hat{U}$ such that in the standard-form, $(\bra{G}_a\otimes \I_s)\hat{Q}(\ket{G}_a\otimes \I_s)=f[\hat{H}]$ for some target operator $f[\cdot]$. Though this is accomplished in prior art using the linear-combination-of-unitaries algorithm~\cite{Kothari2014efficient}, that approach requires a case-by-case detailed analysis of $f$ to obtain the $L_1$ norm of its coefficients in a Taylor expansion, and has a success probability that decays with the inverse square of this norm. 

Our quantum signal processor computes $f[\hat{H}]$ with no such restrictions and with an optimal query complexity that exactly matches polynomial lower bounds for a large class of functions. We call this `quantum signal processing', which generalizes our previous results~\cite{Low2016HamSim} for $d$-sparse oracles to the standard-form and a larger class of functions.
Thus generic improvements to all applications in~\cref{Tab:Applications} can be expected in query complexity, ancilla overhead, and scope of possible signal inputs. In particular,~\cref{thm:QueryComplexity} follows directly from the query complexity of the choice $f(\lambda)=e^{-i\lambda t}$, which corresponds to applying $-t\sin{(\cdot)}$ on eigenphases of the iterate. 

In~\cref{Sec:Quantum_Signal_Processor}, we provide a detailed overview of the quantum signal processor, which uses the standard-form encoding of a matrix $\hat{H}$ in~\cref{def:standardform} as an input. Our claim to the generality of this encoding is justified in~\cref{sec:standardform}, where we map a variety of other common input models describing matrices to the standard-form. The next key result is qubitization, which converts a standard-form encoding to a Grover-like iterate, is proven in~\cref{Sec:Qubitization}. By creatively applying this iterate, we show in~\cref{sec:QSP_without_control}  how a large class of polynomials of $\hat{H}$ may be efficiently computed. Our final result on Hamiltonian simulation by qubitization is proven in~\cref{Sec:Hamiltonian_Simulation} by a simple choice for this polynomial. Though we focus on qubitization of Hermitian matrices in the body of this paper, we also describe the extension to normal matrices in~\cref{Sec:NormalOperators}.

\begin{table}[t]
	\centering
	\begin{tabular}{c||c|c|c|c|c|c}
		Problem & BCCKS~\cite{Berry2015Hamiltonian} & $d$-sparse~\cite{Low2016HamSim} & Evolution by $\rho$ & QPE & QLSP~\cite{Childs2015LinearSystems} & Gibbs~\cite{Chowdhury2016quantum} \\
		\hline
		$\hat{H}$ & Hamiltonian & Hamiltonian & Density matrix  & Unitary & Matrix & Hamiltonian \\
		$\hat{U}$ & Selects $\hat{U}_j$ & Isometry $\hat{T}$ & SWAP  & Any & Any & Any \\
		$\hat{G}$ & $\hat{U}_j$ coefficients & Identity & Purified $\rho$ & Any & Any & Any\\
		\hline
		Solution & $e^{-i\hat{H}t}$ & $e^{-i\hat{H}t}$ & $e^{-i\hat{\rho}t}$ & Decision & $\hat{H}^{-1}$ & $e^{-\beta \hat{H}}$
	\end{tabular}
	\caption{\label{Tab:Applications} 
		List of six example problems (top row), solvable using the quantum signal processor approach to compute an operator function $f[\cdot]$ of Hermitian $\hat{H}=\bra{G}\hat{U}\ket{G}$.
		Through qubitization, the scope of inputs to the Quantum Linear Systems Problem (QLSP) and Gibbs Sampling (Gibbs) can be any $\hat{H}$ of this form, either indirectly through Hamiltonian simulation, or directly through quantum signal processing. Quantum Phase Estimation (QPE) here decides whether eigenphases $\theta$ of an implemented unitary satisfy some property e.g. $f(\theta)\ge 1/2$.
	}
\end{table}

\section{Overview of the Quantum Signal Processor}
\label{Sec:Quantum_Signal_Processor}
Since coherent quantum computation is restricted to unitary operations, one commonly finds a situation such as in~\cref{Tab:Applications}, where post-selection is required to accomplish some desired quantum state transformation. Consider some arbitrary input system quantum state $\ket{\psi}_s\in \mathcal{H}_s$ and suppose that it is transformed by some desired \emph{signal operator} $\hat{H}$ like $\ket{\psi}_s\rightarrow\hat{H}\ket{\psi}_s$. In order to realize this operation, which is non-unitary in general, it is necessary to embed $\hat{H}$ in a larger Hilbert space. 

This is embedding accomplished through some given unitary \emph{signal oracle} $\hat{U}:\mathcal{H}_a\otimes\mathcal{H}_s\rightarrow \mathcal{H}_a\otimes\mathcal{H}_s$ that acts jointly on $\mathcal{H}_s$ and another Hilbert space $\mathcal{H}_a$. The signal oracle encodes $\hat{H}=(\bra{G}_a\otimes\I_s)\hat{U}(\ket{G}_a\otimes\I_s)$ in a subspace flagged by an ancilla \emph{signal state} $\ket{G}_a\in \mathcal{H}_a$. This is the standard-form of~\cref{def:standardform}, which applies $\hat{H}$ as follows.
\begin{align}
\label{Eq:HermitianTransformation}
\hat{U}\ket{G}_a\ket{\psi}_s=\ket{G}_a\hat{H}\ket{\psi}_s+\sqrt{1-\|\hat{H}\ket{\psi}\|^2}\ket{G_\psi^{\perp}}_{as}
,
\quad
\hat{U}=\left(\begin{matrix}
\hat{H} & \cdot \\
\cdot & \cdot
\end{matrix}\right),
\quad 
(\bra{G}_a\otimes\I_s)\ket{G_\psi^{\perp}}_{as}=0.
%\hat{U}&=\left(\begin{matrix}
%\hat{H} & \cdot & \cdot \\
%\sqrt{1-|\hat{H}|^2} & \cdot & \cdot \\
%\cdot & \cdot & \cdot
%\end{matrix}\right),
\end{align}

The signal state defines the measurement basis of the appended register $\mathcal{H}_a$, which naturally divides $\hat{U}$ into two subspaces. First, $\mathcal{H}_G=\ket{G}\otimes\mathcal{H}_s$ where the measurement succeeds with probability $\|\hat{H}\ket{\psi}\|^2$ and $\hat{U}\ket{G}_a\ket{\psi}_s$ is projected onto $\frac{\ket{G}_a\hat{H}\ket{\psi}_s}{\|\hat{H}\ket{\psi}\|}$. Second, the orthogonal complement $\mathcal{H}_{G^\perp}$ where the measurement fails. As probabilities are bounded by $1$, the signal operator $\hat{H}$ must have spectral norm $\|\hat{H}\|\le 1$. Whenever the context is clear, we drop the ancilla and system subscripts, and use $\ket{G}_a\otimes \I_s$ and $\ket{G}$ interchangeably. We represent $\hat{U}$ such that the top-left block is precisely $\hat{H}$ and acts on an input state $\ket{G_\psi}\equiv\ket{G}\ket{\psi}\in\mathcal{H}_G$, whereas the undefined parts of $\hat{U}$ transform $\ket{G_\psi}$ into some orthogonal state $\ket{G_\psi^{\perp}}\in\mathcal{H}_{G^\perp}$ of lesser interest. 

In the following, we consider the case where $\hat{H}$ is a Hermitian matrix. The case of normal $\hat{H}$ is discussed in~\cref{Sec:NormalOperators}. Thus the action of $\hat{U}$ on $\ket{G_\psi}$ in~\cref{Eq:HermitianTransformation} can be written more clearly in the eigenbasis of  $\hat{H}\ket{\lambda}=\lambda \ket{\lambda}$. For each eigenstate of $\hat{H}$,
\begin{align}
\label{Eq:HermitianTransformationEigenstate}
\hat{U}\ket{G}\ket{\lambda}&=\hat{U}\ket{G_\lambda}=\lambda \ket{G_\lambda}+\sqrt{1-|\lambda|^2}\ket{G_\lambda^{\perp}},
\end{align}
and we find it convenient to define the subspace $\mathcal{H}_{\lambda}=\text{span}\{\ket{G_\lambda},\hat{U}\ket{G_\lambda}\}$.
Note the trivial case $\lambda=1$ where $\mathcal{H}_\lambda$ is one-dimensional, which we will ignore. We also emphasize that the action of $\hat{U}$ on states orthogonal to $\ket{G_\lambda}$ are generally not easily controlled by the user, and thus left undefined.

Given any scalar function $f$, our goal is to find an optimal quantum circuit that transforms any standard-form encoding of $\hat{H}$ into a standard-form encoding of $f[\hat{H}]\equiv\sum_{\lambda}f(\lambda)\ket{\lambda}\bra{\lambda}$. We accomplish this with the quantum signal processor.
\\
\begin{definition}[Quantum Signal Processor]
	\label{def:quantumsignalprocessor}
A quantum signal processor solves the following.\\
\begin{tabularx}{\textwidth}{rX}
	Inputs: & \textbullet\;	A Hermitian matrix $\hat{H}$ with bounded spectral norm $\|\hat{H}\|\le1$.
	\\
	& \textbullet\; A function $f:[-1,1]\rightarrow\mathbb{D}$, where $\mathbb{D}$ is the complex unit disc, i.e. $\forall x \in[-1,1],\;|f(x)|\le 1$.
	\\
	\myvspace
	Resources: 
	&
	\textbullet\; An encoding of $\hat{H}=\bra{G}\hat{U}\ket{G}$ by the oracles of~\cref{def:standardform}.
	\\
	&\textbullet\;  A constant number of additional qubits.
	\\
	&\textbullet\;  Arbitrary single and two-qubit gates.
	\\
    \myvspace
	Outputs: 
	& \textbullet\; A standard-form encoding of $f[\hat{H}]$, i.e. a unitary $\hat{Q}$ where $f[\hat{H}]=\bra{0}\hat{Q}\ket{0}$.
\end{tabularx}
\end{definition}

In the simplest case, one might wish to apply $\hat{H}$ multiple times to generate higher moments. For instance, $\hat{H}^2$ would allow a direct estimate of variance. Unfortunately, the subspace $\mathcal{H}_{\lambda}$ for each eigenstate $\ket{\lambda}$ is not invariant under $\hat{U}$ in general. As a result, repeated applications in this basis do not produce higher moments of $\hat{H}$ due to leakage out of $\mathcal{H}_{\lambda}$. The structure of this leakage depends on the undefined components of $\hat{U}$, must be analyzed on a case-by-case basis, and thus is of limited utility.

Order can be restored to this undefined behavior by stemming the leakage. The simplest possibility that preserves the signal operator of~\cref{Eq:HermitianTransformation} replaces $\hat{U}$ with a unitary, the \emph{iterate} $\hat{W}$, that also encodes $\hat{H}=\bra{G}\hat{W}\ket{G}$, but for each eigenstate $\ket{\lambda}$, performs a rotation in SU$(2)$ on disjoint two-dimensional subspaces $\mathcal{H}_{\lambda}=\text{span}\{\ket{G_\lambda},\hat{W}\ket{G_\lambda}\}=\text{span}\{\ket{G_\lambda},\ket{G^{\perp}_\lambda}\}$. This defines the state $\ket{G^{\perp}_\lambda}$ through Gram-Schmidt orthogonalization, and a set of Pauli operators $\hat{X}_\lambda,\hat{Y}_\lambda,\hat{Z}_\lambda$ for each subspace.
\begin{align}
\ket{G^{\perp}_\lambda}=\frac{(\hat{W}-\lambda)\ket{G_\lambda}}{\sqrt{1-|\lambda|^2}},
\quad
\hat{X}_\lambda\ket{G_\lambda}=\ket{G^{\perp}_\lambda},
\quad
\hat{Y}_\lambda\ket{G_\lambda}=i\ket{G^{\perp}_\lambda},
\quad
\hat{Z}_\lambda\ket{G_\lambda}=\ket{G_\lambda}.
\end{align}
In this basis, the iterate for each eigenvalue of $\hat{H}$ is defined to be exactly
\begin{align}
\label{Eq.SubMatrix}
\hat{W}=\left(\begin{matrix}
\lambda & -\sqrt{1-|\lambda|^2} \\
\sqrt{1-|\lambda|^2} & \lambda 
\end{matrix}\right)_\lambda= 
\begin{array}{rr}
\lambda\ket{G_\lambda}\bra{G_\lambda}
&-\sqrt{1-|\lambda|^2}\ket{G_\lambda}\bra{G^{\perp}_\lambda}\\
+\sqrt{1-|\lambda|^2}\ket{G^{\perp}_\lambda}\bra{G_\lambda}
&+\lambda\ket{G^{\perp}_\lambda}\bra{G^{\perp}_\lambda}
\end{array},
\end{align}
On a general input  $\ket{G}\sum_\lambda a_\lambda\ket{\lambda}\in\mathcal{H}_G$, the iterate is represented by
\begin{align}
\label{Eq:W_Block_form}
\hat{W}=
\bigoplus_{\lambda}\left(\begin{matrix}
\lambda & -\sqrt{1-|\lambda|^2}. \\
\sqrt{1-|\lambda|^2}. & \lambda
\end{matrix}\right)_\lambda
=
\bigoplus_{\lambda}
e^{-i\hat{Y}_\lambda \theta_\lambda },
%.
%\bra{G}\hat{W}\ket{G}=\hat{H}.
\end{align}
where $\theta_\lambda$ is defined through the equality
\begin{align}
\label{eq:iterate_angle}
\forall \lambda\in[-1,1],\;\lambda-i\sqrt{1-|\lambda|^2}=\cos{(\theta_\lambda)}-i\sin{(\theta_\lambda)}
\Rightarrow
\theta_\lambda = \cos^{-1}{(\lambda)} .
\end{align}

In the following, the iterate will always be applied to states spanned by the subspace $\bigoplus_\lambda \mathcal{H}_{\lambda}$. Thus its action on states outside this subspace need not be defined. The usefulness of this construct is evident. Due to its invariant subspace, multiple applications of the iterate result in highly structured behavior. However, implementing $\hat{W}$ appears difficult in general. One one hand, the function $\sqrt{1-|\lambda|^2}$ must be computed for each eigenstate. In principle, this can be approximated using phase estimation with considerable overhead~\cite{Daskin2016ancilla}. On the other hand, it is not at all clear whether a unitary with the form of $\hat{W}$ can always be engineered from the standard-form encoding of $\hat{H}$. `Qubitization' is our solution to constructing this iterate with minimal overhead.
\begin{theorem}[Qubitization]
	\label{thm:qubitization}
	Given a Hermitian matrix $\hat{H}=\bra{G}_a\hat{U}\ket{G}_a$ encoded in standard-form as described in~\cref{def:standardform}, the iterate $\hat{W}$ of~\cref{Eq:W_Block_form} can be constructed using at most one query each to $\hat{G}$, controlled-$U$, their inverses, at most one additional qubit, and $\mathcal{O}(\log{(\dim{(\mathcal{H}_a)})})$ quantum gates.
\end{theorem}

Using the oracles $\hat{G},\hat{U}$, and arbitrary unitary operations on \emph{only} the ancilla register, we also provide necessary and sufficient conditions in~\cref{Thm:Subspace}, for when $\hat{W}$ can be implemented exactly using only one query to $\hat{U}$. As these conditions are somewhat restrictive, we then prove in~\cref{Thm.Existence_of_S} that qubitization is \emph{unconditionally} possible by instead using the \emph{controlled-}$\hat{U}$ oracle in a quantum circuit that generates the same $\hat{H}$ and satisfies these conditions. We describe a similar construction for normal operators in Appendix.~\ref{Sec:NormalOperators}.

Observe that $\hat{W}^N$ efficiently produces Chebyshev polynomials $T_N[\hat{H}]$~\cite{Childs2015LinearSystems}. We call any function $[\cdot]$ of the signal $\hat{H}$ \emph{target operators} when they occur in the top-left block and are thus automatically in standard-form. The fact that Chebyshev polynomials are the best polynomial basis for $L_\infty$ function approximation on an finite interval~\cite{Meinardus1967} suggests that the any target operator $f[\hat{H}]=A[\hat{H}]+i B[\hat{H}]$ could be approximated with a judicious choice of controls on the ancilla register. 
%As all Hermitian operators are also quantum observables, achieving this would be of great utility to designing measurements on quantum states.
% - the other essential part of the quantum computing toolbox. 
We present two implementation of the quantum signal processor described in~\cref{def:quantumsignalprocessor}. The first does not require any additional qubits beyond that for qubitization, and realizes a broad class of target operators $f$.
\begin{theorem}[Ancilla-Free Quantum Signal Processing] A quantum signal processor can be implemented with the following properties.
	\\
	\label{thm:AchievableAB}
	\begin{tabularx}{\textwidth}{rX}
		Inputs 
		& \textbullet\;	The iterate $\hat{W}$ obtained from the qubitization procedure of~\cref{thm:qubitization} on a Hermitian matrix $\hat{H}=\bra{G}\hat{U}\ket{G}$ encoded by the oracles of~\cref{def:standardform}.
		\\
		& \textbullet\; Real polynomials $A(\lambda), B(\lambda)$ of degree $Q$ and equal parity satisfying all following.
		\\
		&\qquad \textbullet\;$A(1)=1$;
		\\
		&\qquad \textbullet\;$\forall \lambda \in[-1,1],\; A^2(\lambda)+B^2(\lambda)\le 1$;
		\\
		&\qquad \textbullet\;$\forall \lambda\ge 1,\; A^2(\lambda)+B^2(\lambda)\ge 1$;
		\\
		&\qquad \textbullet\;$\forall Q\text{ even}, \lambda\ge 0,\; A^2(i\lambda)+B^2(i\lambda)\ge 1$.
		%\\
		\\
		\myvspace
		Output:
		&
		\textbullet\; A standard-form encoding of $A[\hat{H}]+iB[\hat{H}]$.
		\\
		\myvspace
		Cost
		& \textbullet\; $Q$ queries to $\hat{W}$.
		\\
		& \textbullet\; Zero additional qubits beyond that for $\hat{W}$.
		\\
		& \textbullet\; $\mathcal{O}(Q\log{(\dim{(\mathcal{H}_a)})})$ arbitrary single and two-qubit gates.
	\end{tabularx}
\end{theorem}
	
Note that the two polynomials in our solution of ancilla-free quantum signal processing are of the same parity. At this point, we assume that all $\hat{U}$ have been qubitized and so use the iterate $\hat{W}$ as our basic building block. A different basis set of functions with fewer restrictions on parity can be obtained by embedding $\hat{W}$ into yet another $\text{SU}(2)$ invariant subspace by adding an ancilla qubit. This leads to the second implementation of the quantum signal processor of~\cref{def:quantumsignalprocessor}.
\begin{theorem}[Single-Ancilla Quantum Signal Processing] A quantum signal processor can be implemented with the following properties.
	\\
	\label{thm:AchievableAC}
	\begin{tabularx}{\textwidth}{rX}
		Inputs: 
		& \textbullet\;	The iterate $\hat{W}$ obtained from the qubitization procedure of~\cref{thm:qubitization} on a Hermitian matrix $\hat{H}=\bra{G}\hat{U}\ket{G}$ encoded by the oracles of~\cref{def:standardform}.
		\\
		& \textbullet\; Real polynomials $A(\lambda), C(\lambda)$ of degree $Q/2$ and opposite parity satisfying all following.
		\\
		&\qquad \textbullet\;$A(0)=1$;
		\\
		&\qquad \textbullet\;$\forall \lambda \in[-1,1],\; A^2(\lambda)+C^2(\lambda)\le 1$;
		%\\
		\\
		\myvspace
		Output: 
		&
		\textbullet\;  A standard-form encoding of $A[\hat{H}]+iC[\hat{H}]$.
		\\
		\myvspace
		Cost: 
		& \textbullet\; $Q$ queries to controlled-$\hat{W}$.
		\\
		& \textbullet\; One additional qubit beyond that for $\hat{W}$.
		\\
		& \textbullet\; $\mathcal{O}(Q\log{(\dim{(\mathcal{H}_a)})})$ arbitrary single and two-qubit gates.
	\end{tabularx}
\end{theorem}

These powerful tools for target operator processing, made possible by qubitization, are agnostic to the underlying oracles that describe the signal operator $\hat{H}$. In many instances, converting this  description to the standard-form of~\cref{def:standardform} is straightforward  and is indeed how our Hamiltonian simulation results for the varied input models of~\cref{Tab:Comparison} are proven. 
%The linear-combination-of-unitaries algorithm~\cite{Kothari2014efficient}, reviewed in~\cref{Sec:LCU}[maybe appendix?], is one such case.
%As such, we conjecture that this standard-form is most natural for quantum algorithms involving matrices, and furnishes an optimal approach to implementing arbitrary quantum measurements and their operator transformations when combined with our quantum signal processing algorithms. This is supported by the fact that the optimal query complexity of our Hamiltonian simulation algorithm stems directly from how quantum signal processing achieves fundamental lower bounds in polynomial approximation theory.

% in implementing arbitrary operator transformations.

%Having motivated our perspective on the problem of qubitization in a quantum signal processor and its significance to designing arbitrary operator transformations, we present the solution.

%of the qubitization can be made fully constructive with an approach for implementing some $\hat{G},\hat{U}$ that encode any desired $\hat{H}$.
\section{Explicit Encodings of $\hat{H}=(\langle G|\otimes\I)\hat{U}(|G\rangle\otimes\I)$}
\label{sec:standardform}
We now justify our motivation for encoding matrices $\hat{H}$ in the standard-form format of $\hat{H}=\bra{G}\hat{U}\ket{G}$ in~\cref{def:standardform}. A number of common techniques for encoding matrix problems on quantum computers map naturally to the standard-form with minimal overhead. Thus taking the standard-form as our starting point is without any loss of generality. This is demonstrated by the following three explicit implementation of the oracles $\hat{U}$ and $\hat{G}\ket{0}=\ket{G}$ for Hamiltonians represented as a linear combination of unitaries, $d$-sparse Hamiltonians, and a new input model where Hamiltonians are represented by a purified density matrix.

\subsection{Linear Combination of Unitaries}
One option for implementing the standard-form encoding is provided by the Linear-Combination-of-Unitaries algorithm (LCU)~\cite{Childs2012,Kothari2014efficient}, which underlies the BCCKS simulation algorithm. LCU is based on the fact that any complex $\hat{H}$ is a linear combination of some $d$ unitary operators:
\begin{align}
\label{Eq:H_decomposition}
\hat{H}=\sum_{j=1}^d \alpha_j\hat{U}_j,\quad \|\hat{H}\| \le \|\vec{\alpha}\|_1= \sum^d_{j=1}|\alpha_j|,
\end{align}
where the upper bound on the spectral norm is $\|\vec{\alpha}\|_1$. Note that this bound depends on the choice of decomposition, but is tight for the some choice. Without loss of generality, all $\alpha_j\ge0$ by absorbing complex phases into $\hat{U}_j$.
The algorithm assumes that the $\alpha_j$ are provided as a list of $d$ numbers, and each $\hat{U}_j$ is provided as a quantum circuit composed of constant number $\mathcal{O}(C)$ of primitive gates. With these inputs, the oracles
\begin{align}
\label{Eq:AV_definition}
\hat{G}=\sum^d_{j=1}\sqrt{\frac{\alpha_j}{\|\vec{\alpha}\|_1}}\ket{j}\bra{0}_a+\cdots ,\quad
\hat{U}=\sum_{j=1}^d\ket{j}\bra{j}_a\otimes\hat{U}_j, \quad \bra{G}\hat{U}\ket{G}=\frac{\hat{H}}{\|\vec{\alpha}\|_1}
\end{align}
can be constructed, where the ancilla state creation operator $\hat{G}\ket{0}=\ket{G}$ is implemented with $\mathcal{O}(d)$ primitive gates, 
%as only is action on one computational basis state is defined,
and the selector $\hat{U}$ is implemented with $\mathcal{O}(dC)$ primitive gate. By direct expansion of $\hat{U}\hat{G}\ket{0}_a\ket{\psi}_s$, this leads exactly to~\cref{Eq:HermitianTransformation}. This proves the following.
\begin{lemma}[Standard-Form Encoding by a Linear Combination of Unitaries]
	\label{cor:encoding_LCU}
	Let $\hat{G}$ prepare the state $\ket{G}_a=\sum^d_{j=1}\sqrt{\alpha_j/\|\vec{\alpha}\|_1}\ket{j}_a$ where $\alpha_j\ge 0$. Let
	$\hat{U}=\sum_{j=i}^d\ket{j}\bra{j}_a\otimes\hat{U}_j$. These oracles encode the matrix
$
	\bra{G}\hat{U}\ket{G}=\frac{1}{\|\vec{\alpha}\|_1}\sum_{j=1}^d\alpha_j \hat{U}_j
$.
\end{lemma}
\begin{proof}
	Consider the computation
	\begin{align}
	(\bra{G}_a\otimes\I_s)\hat{U}(\ket{G}_a\otimes\I_s)
	&=(\bra{G}_a\otimes\I_s)\left(\sum^d_{j=1}\sqrt{\frac{\alpha_j}{\|\vec{\alpha}\|_1}}\ket{j}_a\otimes \hat{U}_j\right)
	=\sum^d_{j,k=1}\frac{\sqrt{\alpha_j\alpha_k}}{\|\vec{\alpha}\|_1}\stab{k|j}\otimes \hat{U}_j
	\nonumber\\
	&=\frac{1}{\|\vec{\alpha}\|_1}\sum_{j=1}^d\alpha_j \hat{U}_j.
	\end{align}
\end{proof}

Of course, the optimal decomposition that costs the fewest number of ancilla qubits and primitive gates may be difficult to find, and may not even fit naturally in this model, but LCU shows that implementing an encoding for any $\hat{H}$ is possible in principle.

\subsection{$d$-Sparse Hamiltonians}
The model of $d$-sparse Hamiltonians is another paradigm for specifying matrices to quantum computers, and is particularly common in the design of quantum algorithms by quantum walks~\cite{Aharonov2003Adiabatic}. Such Hamiltonians have at most $d$ non-zero entries in any row, and information on their positions and matrix values $\hat{H}_{jk}$ are accessed through the following two standard unitary oracles~\cite{Berry2012}.
\begin{align}
\label{eq:oracles_sparse}
\hat{O}_{H}\ket{j}\ket{k}\ket{z}=\ket{j}\ket{k}\ket{z\oplus \hat{H}_{jk}},
\quad
\hat{O}_{F}\ket{j}\ket{l}=\ket{j}\ket{f(j,l)}.
\end{align}
Observe that $\hat{O}_H$ accepts a row $j$ and column $k$ index and returns $\hat{H}_{jk}$ in some binary format. The other oracle $\hat{O}_F$ accepts a row $j$ index and a number $l\in[d]$ to compute in-place the column index $f(j,l)$ of the $l^\text{th}$ non-zero element in row $j$. Given this input, we now encode $\hat{H}$ in standard-form. 
\begin{lemma}[Standard-Form Encoding of a $d$-Sparse Hamiltonian]
	\label{cor:encoding_sparse}
	Let the oracles of~\cref{eq:oracles_sparse} specify a $d$-sparse Hamiltonian $\hat{H}$ with max-norm $\|H\|_\text{max}$. Then the oracles encoding $\bra{G}\hat{U}\ket{G}=\frac{\hat{H}}{d\|\hat{H}\|_{\text{max}}}$ can be implemented using $\mathcal{O}(1)$ queries.
\end{lemma}
\begin{proof}
	Let $\ket{G}=\ket{0}_{a_1}\ket{0}_{a_2}\ket{0}_{a_3}\equiv\ket{0}_a$ be a computational basis state, and let $F_j=\{f(j,l)\}_{l\in[d]}$ be the set of column indices to all non-zero elements in row $j$. Ref.~\cite{Berry2012} shows how each of the isometries
	\begin{align}
	\hat{T}_1&=\sum_{j}\ket{\psi_j}\bra{0}_a\bra{j}_s,
	\; \ket{\psi_j}=\sum_{p\in F_j}\frac{\ket{p}_{a_3}}{\sqrt{d}}\Big(\sqrt{\frac{H_{pj}}{\|\hat{H}\|_{\text{max}}}}\ket{0}_{a_1}+\sqrt{1-\frac{|H_{pj}|}{\|\hat{H}\|_{\text{max}}}}\ket{1}_{a_1}\Big)\ket{0}_{a_2}\ket{j}_s,	
	\\ \hat{T}_2&=\sum_{k}\ket{\chi_k}\bra{0}_a\bra{k}_s,
	\; 
	\bra{\chi_k}=\sum_{p\in F_k}\frac{\bra{p}_{s}}{\sqrt{d}}\Big(\sqrt{\frac{H_{kp}}{\|\hat{H}\|_{\text{max}}}}\bra{0}_{a_2}+\sqrt{1-\frac{|H_{kp}|}{\|\hat{H}\|_{\text{max}}}}\bra{1}_{a_2}\Big)\bra{0}_{a_1}\bra{k}_{a_3},
	\nonumber
	\end{align} 
	can be implemented using using $2$ queries to $\hat{O}_H$ and $1$ query to $\hat{O}_F$. By construction, the overlap of these states is $
	\stab{\chi_k|\psi_j}=\frac{H_{kj}}{d\|\hat{H}\|_{\text{max}}}$. Now choose $\hat{U}=\hat{T}_2^{\dag}\hat{T}_1$. By a direct computation,
	\begin{align}
(\bra{G}_a\otimes\I_s)\hat{U}(\ket{G}_a\otimes\I_s)
&=\sum_{k,j}\ket{k}_s	\stab{\chi_k|\psi_j}\bra{j}_s=\sum_{k,j}\frac{H_{kj}}{d\|\hat{H}\|_{\text{max}}}\ket{k}\bra{j}_s=\frac{\hat{H}}{d\|\hat{H}\|_{\text{max}}}.
	\end{align}
\end{proof}

\subsection{Purified Density Matrix}
The simplicity of the standard-form encoding allows us to swiftly devise new input models for Hamiltonians. Here, we consider the case where a Hamiltonian $\hat{H}=\tr{\ket{G}\bra{G}}_{a1}$ is a density matrix $\hat{\rho}$ obtained by tracing out the ancilla register of a pure state prepared by some oracle $\hat{G}$. That is,
\begin{align}
\label{eq:oracle_density_matrix}
\hat{G}\ket{0}_a=\ket{G}_a=\sum_{j}\sqrt{\alpha_j}\ket{j}_{a_1}\ket{\chi_j}_{a_2},
\quad
\hat\rho = \tr{\ket{G}\bra{G}}_{a1}=\sum_{j}\alpha_j\ket{\chi_j}\bra{\chi_j}.
\end{align}
\begin{lemma}[Standard-Form Encoding of a Purified Density Matrix]
	\label{cor:encoding_purified}
	Let the oracle of~\cref{eq:oracle_density_matrix} prepare a state $\hat{G}\ket{0}_a=\ket{G}_a=\sum_{j}\sqrt{\alpha_j}\ket{j}_{a_1}\ket{\chi_j}_{a_2}$ that is a purification of the density matrix $\hat{\rho}=\sum_{j}\alpha_j\ket{\chi_j}\bra{\chi_j}=\tr{\ket{G}\bra{G}}_{a_1}$. Let $\hat{U}$ be a unitary that swaps the register $a_2$ with the system register $s$. These oracles encode the matrix
	$
	\bra{G}\hat{U}\ket{G}=\hat{\rho}
	$.
\end{lemma}
\begin{proof}
	Let $\{\ket{\lambda}\}$ be a complete basis on the system. By a direct computation,
	\begin{align}
	\bra{G}\hat{U}\ket{G}_a \sum_{\lambda}\ket{\lambda}\bra{\lambda}_s
	=\sum_{\lambda}\sum_{j}\bra{G}\sqrt{\alpha_j}\ket{j}_{a_1}\ket{\lambda}_{a_2}\ket{\chi_j}\bra{\lambda}_s
	=\sum_{\lambda}\sum_{j}|\alpha_j| \ket{\chi_j}_s\langle \chi_j|\lambda\rangle\bra{\lambda}_s = \hat{\rho}.
	\end{align}
\end{proof}

\section{Qubitization in a Quantum Signal Processor}
\label{Sec:Qubitization}
This section describes qubitization in detail: the process for creating the iterate $\hat{W}$ given $\hat{H}$ encoded in standard-form, and an essential component in a systematic procedure for implementing operator transformations of $\hat{H}$. In~\cref{Thm:Subspace}, we provide necessary and sufficient conditions on when $\hat{W}$ can be constructed from the oracles $\hat{G},\hat{U}$. Then in~\cref{Thm.Existence_of_S}, we show that any $\hat{G},\hat{U}$ not satisfying these conditions can be efficiently transformed into a $\hat{G}',\hat{U}'$ that do, and encode the same signal operator $\hat{H}=\bra{G'}\hat{U}'\ket{G'}$. Together, these Lemmas in the remainder of this section complete the proof of~\cref{thm:qubitization} for the case of Hermitian $\hat{H}$.

%A constructive approach to implementing an encoding $\hat{U}$ for any desired $\hat{H}$ is then reviewed in~\cref{Sec:LCU}.

For now, we assume that $\ket{G}$ is known. This soon proves to be unnecessary and only oracle access to $\hat{G}$ is required. Thus we must find a unitary $\hat{S}'$, acting only on the ancilla register such that the iterate $\hat{W}=\hat{S}'\hat{U}$ of~\cref{Eq:W_Block_form} is obtained. For the case of Hermitian $\hat{H}\ket{\lambda}=\lambda\ket{\lambda}$, we now determine necessary and sufficient conditions on what $\hat{S}'$ must be. As $\hat{S}'$ is otherwise arbitrary, we use without loss of generality the ansatz of $\hat{S}'$ being a product of a reflection about $\ket{G}$ and another arbitrary unitary $\hat{S}$ on the ancilla:
\begin{align}
\label{Eq:Reflection}
\hat{W}&=((2\ket{G}\bra{G}-\I)_a \otimes \I_s)\hat{S}\hat{U},\quad \ket{G_\lambda}=\ket{G}\ket{\lambda}\Rightarrow \ket{G^{\perp}_{\lambda}}=\frac{\lambda \ket{G_\lambda}-\hat{S}\hat{U}\ket{G_\lambda}}{\sqrt{1-\lambda^2}}.
\end{align}
\begin{lemma}[Conditions on Qubitization]
	\label{Thm:Subspace}
	For all signal oracles $\hat{U}$ that implement the Hermitian signal operator
	% of~\cref{Eq:HermitianTransformation}
	$\hat{H}$, the unitary $\hat{S}$ in~\cref{Eq:Reflection} creates a unitary iterate $\hat{W}$ with the same signal operator in the same basis, but in an $\text{SU}(2)$ invariant subspace containing $\ket{G}$ if and only if
	\begin{align}
	\label{Eq:Necessary_Sufficient_SVSV}
	\bra{G}_a\hat{S}\hat{U}\ket{G}_a= \hat{H}\text{ and }\bra{G}_a\hat{S}\hat{U}\hat{S}\hat{U}\ket{G}_a=\I.
	\end{align}
\end{lemma}
\begin{proof}
	In the forward direction, we assume~\cref{Eq:W_Block_form}, then compute and compare with~\cref{Eq:Reflection}: $\lambda=\bra{G_\lambda}\hat{W}\ket{G_\lambda}=\bra{G_\lambda}\hat{S}\hat{U}\ket{G_{\lambda}}$. By using this result repeatedly together with the fact that $\hat{S}\hat{U}$ is unitary, Gram-Schmidt orthonormalization of $\hat{W}\ket{G_\lambda}$ furnishes the state $\ket{G^{\perp}_{\lambda}}
	=
	\frac{\lambda \ket{G_{\lambda}}-\hat{S}\hat{U}\ket{G_{\lambda}}}{\sqrt{1-\lambda^2}}$ which is orthogonal to $\ket{G_\lambda}$. By similarly computing and comparing
	$-\sqrt{1-\lambda^2} = \bra{G_\lambda}\hat{W}\ket{G^{\perp}_{\lambda}}=\frac{\lambda^2-\bra{G_\lambda}(\hat{S}\hat{U})^2\ket{G_\lambda}}{\sqrt{1-\lambda^2}}$, 
	we obtain
	$\bra{G_\lambda}(\hat{S}\hat{U})^2\ket{G_\lambda} = 1$.
	As these must be true for all eigenvectors $\ket{\lambda}$, the conditions in~\cref{Eq:Necessary_Sufficient_SVSV} are necessary. 
	
	That these are also sufficient follows from assuming~\cref{Eq:Necessary_Sufficient_SVSV} and attempting to recover the components of~\cref{Eq:W_Block_form} using the definitions of~\cref{Eq:Reflection}. By applying 
	$\bra{G}_a\hat{S}\hat{U}\ket{G}_a= \hat{H}$,
	we compute 
	$\hat{W}\ket{G_\lambda}=2\ket{G}_a\bra{G}_a\hat{S}\hat{U}\ket{G_\lambda}-\hat{S}\hat{U}\ket{G_\lambda}=2\lambda\ket{G_\lambda}-\hat{S}\hat{U}\ket{G_\lambda}$.
	In the basis of $\ket{G_\lambda}$ and $\ket{G_\lambda^\perp}$, $\bra{G_\lambda}\hat{W}\ket{G_\lambda}=2\lambda-\bra{G_\lambda}\hat{S}\hat{U}\ket{G_\lambda}=\lambda$ 
	and 
	$\bra{G_\lambda^\perp}\hat{W}\ket{G_\lambda}=\frac{\bra{G_\lambda}\lambda -\bra{G_\lambda}(\hat{S}\hat{U})^\dag}{\sqrt{1-\lambda^2}}(2\lambda\ket{G_\lambda}-\hat{S}\hat{U}\ket{G_\lambda})=\frac{2\lambda^2-2\lambda^2-\lambda^2+1}{\sqrt{1-\lambda^2}}=\sqrt{1-\lambda^2}$. 
	A similar calculation for the remaining components requires $\bra{G}_a\hat{S}\hat{U}\hat{S}\hat{U}\ket{G}_a=\I$ and reveals that $\bra{G_\lambda^\perp}\hat{W}\ket{G_\lambda^\perp}=\lambda$ and $\bra{G_\lambda}\hat{W}\ket{G_\lambda^\perp}=-\sqrt{1-\lambda^2}$. As this must be true for all $\lambda$, we may indeed represent $\hat{W}=\bigoplus_{\lambda}\left(\begin{smallmatrix}
	\lambda  & -\sqrt{1-\lambda^2} \\
	\sqrt{1-\lambda^2} & \lambda 
	\end{smallmatrix}\right)_\lambda$.
\end{proof}

In hindsight, these results are manifest. After all, $\bra{G}\hat{S}\hat{U}\hat{S}\hat{U}\ket{G}=\I$ implies that $\hat{S}\hat{U}$ is a reflection when controlled by input state $\ket{G}$, and it is well-known that a Grover iterate~\cite{grover1996fast,Yoder2014} is the product of two reflection about start and target subspaces. Nevertheless, the sufficiency of these conditions highlights that this is the simplest method to extract controllable and predictable behavior out of $\hat{U}$. In particular, these conditions are automatically satisfied in the trivial case with $\hat{S}=\I_{as}$ when $\hat{U}$ only has eigenvalues $\pm 1$, such as when it is a controlled-Pauli operator. 
\begin{corollary}[Qubitization of Reflections]
	\label{Thm:Subspace2}
	For all signal oracles $\hat{U}$ that satisfy $\hat{U}^2=\hat{I}_{as}$ and implement the Hermitian signal operator $\bra{G}_a\hat{U}\ket{G}_a= \hat{H}$, the unitary iterate $\hat{W}=((2\ket{G}\bra{G}-\I)_a \otimes \I_s)\hat{U}$ implements the same signal operator in the same basis, but in an $\text{SU}(2)$ invariant subspace containing $\ket{G}$.
\end{corollary}
\begin{proof}
	Set $\hat{S}=\I_{as}$ in~\cref{Thm:Subspace}, and verify that~\cref{Eq:Necessary_Sufficient_SVSV} is satisfied.
\end{proof}

Unfortunately, a solution to~\cref{Eq:Necessary_Sufficient_SVSV} may not exist for more general $\hat{U}$.~\cref{Thm:Subspace}, amounts to choosing $\hat{S}$ such that $\hat{S}\hat{U}\hat{S}$ is the inverse $\hat{U}^{\dag}$ whist preserving the signal operator $\bra{G}\hat{S}\hat{U}\ket{G}= \hat{H}$. Given that $\hat{S}$ only acts on the ancilla register, it is hard to see how this is always possible. Even if so, $\hat{S}$ may be difficult to implement as it is an arbitrary unitary acting on a potentially large ancilla register. The solution is to construct a different quantum circuit $\hat{U}'$ that contains $\hat{U}$ but still implements the same signal operator, and crucially always has a extremely simple solution $\hat{S}$. We now show how this can be done in all cases using only $1$ query to controlled-$\hat{U}$ and controlled-$\hat{U}^{\dag}$.  
\begin{lemma}[Existence of Qubitization]
	\label{Thm.Existence_of_S}
	For all signal unitaries $\hat{U}$ that implement the Hermitian signal operator $\bra{G}\hat{U}\ket{G}=\hat{H}$, there exists a quantum circuit $\hat{U}'$, using one more qubit, that queries controlled-$\hat{U}$ and controlled-$\hat{U}^{\dag}$ once to implements the same signal operator. Moreover, $\hat{U}'$ satisfies the conditions~\cref{Eq:Necessary_Sufficient_SVSV}
\end{lemma}
\begin{proof}
	We prove this by an explicit construction. Let the controlled-$\hat{U}$ operators be $\hat{V}_1=\ket{0}\bra{0}\otimes\I+\ket{1}\bra{1}\otimes\hat{U}^{\dag}$, $\hat{V}_2=\ket{0}\bra{0}\otimes\hat{U}+\ket{1}\bra{1}\otimes\I$. Thus the extra qubit states $\ket{0},\ket{1}$, are flags that selects either $\hat{U}_j$ or $\hat{U}_j^{\dag}$. By multiplying, $\hat{U}'=\hat{V}_1\hat{V}_2=\ket{0}\bra{0}\otimes\hat{U}+\ket{1}\bra{1}\otimes\hat{U}^{\dag}$.
	Now consider the ancilla state $\ket{G'}=\frac{1}{\sqrt{2}}(\ket{0}+\ket{1})\ket{G}$, and choose $\hat{S}=(\ket{0}\bra{1}+\ket{1}\bra{0})\otimes \I_{as}$. It is easy to verify that the conditions of~\cref{Eq:Necessary_Sufficient_SVSV} is satisfied.
	\begin{align}
	\bra{G'}\hat{S}\hat{U}'\ket{G'}=\bra{G'}\hat{U}'\ket{G'}=\frac{1}{2}(\hat{H}+\hat{H}^{\dag})=\hat{H}, \quad
	\bra{G'}\hat{S}\hat{U}'\hat{S}\hat{U}'\ket{G'}=\bra{G'}\hat{U}'^{\dag}\hat{U}'\ket{G'}=\I,
	\end{align}
	where we have used the fact that $\hat{S}\ket{G'}=\ket{G'}$ is an eigenstate, and that $\hat{S}$ swaps the $\ket{0},\ket{1}$ ancilla states in $\hat{U}'$, thus transforming it into its inverse.
\end{proof}
Even if we are given $\hat{U}$ for which there is no solution to~\cref{Eq:Necessary_Sufficient_SVSV}, we can always apply~\cref{Thm.Existence_of_S} to construct a $\hat{U}'$ that does with minimal overhead. Furthermore our proof uses no information about the detailed structure of $\ket{G}$. Thus without loss of generality, we can assume that any $\hat{G},\hat{U}$ have already been qubitized. 

\section{Operator Function Design on a Quantum Signal Processor}
\label{sec:QSP_without_control}
The purpose of the quantum signal processor is to transform the signal $\hat{H}$ into any desired target operator $f[\hat{H}]$. 
%Schemes for doing so have been previously described and applied important problems~\cite{Berry2015Truncated,childs2015quantum,Chowdhury2016quantum}, but are somewhat ad-hoc and have to be analyzed on a case-by-case basis, also leading to suboptimal query complexity to $\hat{U}$~\cite{Low2016HamSim}. 
We present a systematic framework that furnishes the optimal complexity and a concrete procedure for almost any $f$ %would be of great utility. We now present such a framework for a large class of target operators, 
and show how an exact connection is made between query complexity and the theory of best function approximations with polynomials~\cite{McClellanParksRabiner1973,Meinardus1967}.
%. due to their use of post-selection~\cite{Berry2014,Berry2015Truncated,childs2015quantum} followed by oblivious amplitude amplification~\cite{Berry2014}, our results realize a highly structured and systemic approach with a deep connection to the theory of function approximation. 

Qubitization in~\cref{Sec:Qubitization} is the essential first step that makes this endeavor plausible, as evidenced by the highly structured behavior of the iterate $\hat{W}$ in~\cref{Eq:W_Block_form}, where for Hermitian $\hat{H}$, multiple applications elegantly generate Chebyshev polynomials $T_L[\hat{H}]$~\cite{Childs2015LinearSystems}. To go further, additional control parameters on $\hat{W}$ are necessary, and in the following, we only consider Hermitian $\hat{H}$.
Thus we introduce the \emph{phased iterate} $\hat{W}_\phi$ with the same invariant subspace as $\hat{W}$, and parameterized by $\phi\in\mathbb{R}$.
\begin{align}
\label{Eq:W_phi_Block_form}
\hat{W}_\phi
=
\bigoplus_{\lambda}\left(\begin{matrix}
\lambda  & -ie^{-i\phi}\sqrt{1-|\lambda|^2} \\
-ie^{i\phi}\sqrt{1-|\lambda|^2} & \lambda
\end{matrix}\right)_\lambda
=
\bigoplus_{\lambda}e^{-i\hat\phi^\lambda  \theta_\lambda}
,
%\quad
%\bra{G}\hat{W}_\phi\ket{G}=\hat{H}.
%=e^{-i\hat\sigma_\phi \otimes\cos^{-1}{[\hat{H}]}}
%,\;\;\hat\sigma_\phi=\cos{(\phi)}\hat{\sigma}_x+\sin{(\phi)}\hat\sigma_y.
\end{align}
where $\hat\phi^\lambda=\cos{(\phi)}\hat{X}_\lambda+\sin{(\phi)}\hat{Y}_\lambda$, and the eigenphase $\theta_\lambda$ is defined similar to~\cref{eq:iterate_angle}.
\begin{lemma}
	\label{Cor:PhasedIterate}
	The phased iterate $\hat{W}_\phi$ in~\cref{Eq:W_phi_Block_form} is equal to $\hat{W}_\phi=\hat{Z}_{\phi+\pi/2}\hat{W}\hat{Z}_{-\phi-\pi/2}$, where $\hat{Z}_{\phi}=((1+e^{-i\phi})\ket{G}\bra{G}-\I)$ is a partial reflection about $\ket{G}$ by angle $\phi\in\mathbb{R}$ and implements a relative phase between the $\ket{G_\lambda}$ and $\ket{G^\perp_\lambda}$ subspaces. In block form,
	\begin{align}
	\label{Eq.ZBlockform}
	\hat{Z}_{\phi}=\bigoplus_{\lambda}
	\left(\begin{matrix}
	e^{-i\phi} & 0 \\
	0 & -1
	\end{matrix}\right)_\lambda.
	\end{align}
\end{lemma}
\begin{proof}
	Let us compute the phase applied to states $\ket{G_\lambda},\ket{G^\perp_\lambda}$: $\hat{Z}_{\phi}\ket{G_\lambda}=((1+e^{-i\phi})-1)\ket{G_\lambda}=e^{-i\phi}\ket{G_\lambda}$ and $\hat{Z}_{\phi}\ket{G_\lambda^\perp}=-\ket{G_\lambda^\perp}$. As this true for all $\lambda$,~\cref{Eq.ZBlockform} follows. Combining the representation of $\hat{W}$ from~\cref{Eq:W_Block_form} with this leads to~\cref{Eq:W_phi_Block_form},
	\begin{align}
	\hat{W}_\phi
	\nonumber
	&=\hat{Z}_{\phi-\pi/2}\hat{W}\hat{Z}_{-\phi+\pi/2}
	=\bigoplus_{\lambda}
	\left(\begin{matrix}
	-i e^{-i\phi} & 0 \\
	0 & -1
	\end{matrix}\right)_\lambda
	\left(\begin{matrix}
	\lambda  & -\sqrt{1-|\lambda|^2} \\
	\sqrt{1-|\lambda|^2} & \lambda 
	\end{matrix}\right)_\lambda
	\left(\begin{matrix}
	i e^{i\phi} & 0 \\
	0 & -1
	\end{matrix}\right)_\lambda.
	\end{align}
\end{proof}

We provide algorithms with different overheads for large classes of transformations. `Ancilla-free quantum signal processing' in~\cref{Sec:Observable_Transformation} implement target operators where the real and imaginary parts have the same parity with respect to $\hat{H}$. Opposite parity is obtained by `single-ancilla quantum signal processing' in~\cref{sec:QSP_with_control}.

\subsection{Ancilla-Free Quantum Signal Processing}
\label{Sec:Observable_Transformation}
Consider a sequence of $Q$ of phased iterates, where the angle $\phi$ defining each may differ.
\begin{align}
\label{Eq:WSequenceDef}
\nonumber\hat{W}_{\vec{\phi}}
&
=\hat{W}_{\phi_Q}\cdots \hat{W}_{\phi_2}\hat{W}_{\phi_1}=\bigoplus_\lambda e^{-i\hat\phi_{Q}^{\lambda}  \theta_\lambda}\cdots e^{-i\phi_{2}^{\lambda}   \theta_\lambda}e^{-i\hat\phi_{1}^{\lambda}   \theta_\lambda},  \quad \hat{\phi}_{j}^{\lambda}= 
\cos{(\phi_j)}\hat{X}_\lambda+\sin{(\phi_j)}\hat{Y}_\lambda.
\\
&=
\bigoplus_\lambda 
\mathcal A(2\theta_\lambda)\I_\lambda +i\mathcal B(2\theta_\lambda)\hat Z_\lambda +i\mathcal C(2\theta_\lambda)\hat{X}_\lambda+i \mathcal D(2\theta_\lambda)\hat{Y}_\lambda.
\end{align}
In each subspace $\mathcal{H}_\lambda$, this is a product of $\text{SU}(2)$ rotations. As such, we may decompose this in the Pauli basis $\I_\lambda,\hat{X}_\lambda,\hat{Y}_\lambda,\hat{Z}_\lambda$ with the real functions $(\mathcal A,\mathcal B,\mathcal C,\mathcal D)$ as coefficients.
As we can only prepare and measure in the basis of the state $\ket{G}_a$, consider the component $\bra{G}_a\hat{W}_{\vec{\phi}}\ket{G}_a=\sum_\lambda (\mathcal A(\theta_\lambda)+i \mathcal B(\theta_\lambda))\ket{\lambda}\bra{\lambda}\equiv A[\hat{H}]+iB[\hat{H}]$. Any choice of phases $\vec{\phi}\in\mathbb{R}^Q$ generates sophisticated interference effects between elements of the sequence, leading to $(\mathcal A,\mathcal B,\mathcal C,\mathcal D)$ with some non-trivial functional dependence on $\hat{H}$. Though the dependence of the output on $\vec{\phi}$ seems hard to intuit, they nevertheless specify a \emph{program} for computing functions of $\hat{H}$, similar to how a list of numbers might specify a polynomial.

To understand~\cref{Eq:WSequenceDef}, it suffices to study the following sequence of single-qubit rotations.
\begin{align}
\label{Eq:QSP_Single_Qubit}
\hat{U}_{\vec\phi}&
=e^{-i\hat{\phi}_Q\theta/2 }\cdots e^{-i\hat\phi_2\theta/2}e^{-i\hat\phi_1\theta/2}
=\mathcal  A(\theta)\I+i\mathcal B(\theta)\hat Z+i\mathcal C(\theta)\hat X+i\mathcal D(\theta)\hat Y,
\end{align}
These functions previously characterized in Ref.~\cite{Low2016}, and found the following.
\begin{lemma}[Achievable $A,B$ -- Thm.~2.3 of Ref.~\cite{Low2016}]
	\label{Lem:AchievableAB}
	For any integer $Q>0$, a choice of functions $\mathcal  A(\theta)\equiv A(x),\mathcal B(\theta) \equiv B(x)$ in~\cref{Eq:QSP_Single_Qubit} is achievable by some $\vec\phi\in\mathbb{R}^{Q}$ if and only if all the following are true:\\
	(1) $A(x)$ and $B(x)$ are real parity-$(Q\mod{2})$ polynomials in $x=\cos{(\theta/2)}$ of degree at most $Q$;
	\\
	(2) $A(1)=1$;
	\\
	(3) $\forall x\in[-1,1]$, $A^2(x)+B^2(x)\le 1$;
	\\
	(4) $\forall x\ge 1$, $A^2(x)+B^2(x)\ge 1$;
	\\
	(5) $\forall Q\;\text{even}, x\ge 0$, $A^2(ix)+B^2(ix)\ge 1$.
	\\
	Moreover, $\vec\phi\in\mathbb{R}^{Q}$ can be computed in classical $\mathcal{O}(\text{poly}(Q))$ time.
\end{lemma}
In other words, we may specify only the components $\mathcal A,\mathcal B$, independent of the others $\mathcal C,\mathcal D$ that are of lesser interest. By mapping this result to~\cref{Eq:WSequenceDef}, we complete the proof of~\cref{thm:AchievableAB}.
\begin{proof}[Proof of~\cref{thm:AchievableAB}]
	The unitary operators in each subspace $\mathcal{H}_\lambda$ of $\hat{W}_{\vec\phi}$ are isomorphic to the  product of single qubit rotations $e^{-i\hat{\phi}_Q\theta/2}\cdots e^{-i\hat\phi_2\theta/2}e^{-i\hat\phi_1\theta/2}$. We identify $\theta/2=\theta_\lambda$, where 	$\theta_\lambda=\cos^{-1}(\lambda)$ as defined in~\cref{eq:iterate_angle}.	Thus the result follows from~\cref{Lem:AchievableAB} by substituting $x=\cos{(\theta_\lambda)}=\lambda$.
\end{proof}
As polynomials form a complete basis on bounded real intervals, these results imply the query complexity of approximating any real function $A[\hat{H}]$ with error $\epsilon$ is exactly that of its best polynomial $\epsilon$-approximation satisfying the constraints of~\cref{thm:AchievableAB}, and similarly for the complex case.
% For instance, when $A[\lambda]$ 

\subsection{Single-Ancilla Quantum Signal Processing}
\label{sec:QSP_with_control}
Using no ancilla,~\cref{thm:AchievableAB} implements a target operator $f[\hat{H}]$ where the real and complex parts are constrained to have the same parity. By introducing additional ancilla, other general classes of functions can be implemented, sometimes with looser constraints.
% such as complementary behavior where these parts of the target operator have opposite parity. 
In this section, we use a key observation from Ref.~\cite{Low2016HamSim}. Given any unitary $\hat{V}$ with eigenstates $\hat{V}\ket{\lambda}=e^{i\theta_\lambda}\ket{\lambda}$ and $\hat{V}_0=\ket{+}\bra{+}_b\otimes\I_{s}+\ket{-}\bra{-}_b\otimes\hat{V}$ controlled by the single-qubit ancilla register $b$ where $\hat X_b\ket{\pm}_b=\pm\ket{\pm}_b$, consider the sequence 
\begin{align}
\label{Eq:QSP}
\hat{V}_{\vec{\varphi}}=\prod^{Q/2}_{k\text{ odd}\ge1}\hat{V}^{\dag}_{\varphi_{k+1}+\pi}\hat{V}_{\varphi_k}= \hat{V}^{\dag}_{\varphi_Q+\pi}\cdots \hat{V}_{\varphi_3}\hat{V}^{\dag}_{\varphi_2+\pi}\hat{V}_{\varphi_1},
\;
\hat{V}_\varphi=(e^{-i\varphi\hat{Z}/2}\otimes \I_{s})\hat{V}_0 (e^{i\varphi\hat{Z}/2}\otimes \I_{s}). 
\end{align}
For each eigenstate $\ket{\lambda}$, we obtain a product of single qubit operators $e^{-i \hat{\varphi}_Q\theta_\lambda/2}\cdots e^{-i\hat{\varphi}_1\theta_\lambda/2}$ similar to~\cref{Eq:WSequenceDef} but with a halved phased, and these only act on the ancilla $b$. Using the same reasoning as in~\cref{Sec:Observable_Transformation}, the choice of $\vec{\varphi}$ determines the effective single-qubit ancilla operator
\begin{align}
\label{Eq.QSPcomponents}
\hat{V}_{\vec{\varphi}}=\bigoplus_\lambda\left(\I_b \mathcal A(\theta_\lambda)+i\hat{Z}_b \mathcal  B(\theta_\lambda)+i\hat{X}_b \mathcal  C(\theta_\lambda)+i\hat{Y}_b \mathcal  D(\theta_\lambda)\right)\otimes\ket{\lambda}\bra{\lambda}_s.
\end{align}
In the following, we focus on the functions $\mathcal{A},\mathcal{C}$, which we also characterized fully in previous work.
\begin{lemma}[Achievable $(\mathcal A,\mathcal C)$ -- Thm.~1 of Ref.~\cite{Low2016HamSim}]
	\label{Lem:AchievableAC}
	For any even integer $Q>0$, a choice of functions $\mathcal A(\theta),\mathcal C(\theta)$ in~\cref{Eq:QSP_Single_Qubit} is achievable by some $\vec\phi\in\mathbb{R}^{Q}$ if and only if all the following are true:\\
	(1) $\mathcal A(\theta)=\sum_{k=0}^{Q/2}a_k\cos{(k\theta)}$ is a real cosine Fourier series of degree at most $Q/2$;
	\\
	(2) $\mathcal C(\theta)=\sum_{k=1}^{Q/2}c_k\sin{(k\theta)}$ is a real sine Fourier series of degree at most $Q/2$;
	\\
	(3) $\mathcal{A}(0)=1$;
	\\
	(4) $\forall\theta\in\mathbb{R}$, $\mathcal A^2(\theta)+\mathcal C^2(\theta)\le 1$.
	\\
	Moreover, $\vec\phi\in\mathbb{R}^{Q}$ can be computed in classical $\mathcal{O}(\text{poly}(Q))$ time.
\end{lemma}

In some cases, one might be given target functions $\mathcal A, \mathcal C$ that are only $\epsilon$-close to being achievable, for instance, if $\mathcal A, \mathcal C$ are the output of some numerical procedure. This poses no fundamental difficulty, as we prove in the following, which generalizes slightly a similar statement in Ref.~\cite{Low2016HamSim}. 
\begin{lemma}[Stability of Achievable $(\mathcal A,\mathcal C)$]
	\label{Lem:StabilityAC}
	For any even integer $Q>0$, let
	\\
	(1) $\tilde{ \mathcal A}(\theta)=\sum_{k=0}^{Q/2}a_k\cos{(k\theta)}$ be a real cosine Fourier series of degree at most $Q/2$;
	\\
	(2) $\tilde{ \mathcal C}(\theta)=\sum_{k=1}^{Q/2}c_k\sin{(k\theta)}$ be a real sine Fourier series of degree at most $Q/2$;
	\\
	(3) $\tilde{ \mathcal{A}}(0)=1+\epsilon_1$, where $\epsilon_1\le 1$;
	\\
	(4) $\forall\theta\in\mathbb{R}$, $\tilde{ \mathcal A}^2(\theta)+\tilde{ \mathcal C}^2(\theta)\le 1+\epsilon_2$, where $\epsilon_2\in[0,1]$.
	\\
	Then there exists Fourier series $\mathcal A,\mathcal C$ that satisfy the conditions of~\cref{Lem:AchievableAC}, approximate
	\begin{align} \max_{\theta\in\mathbb{R}}|(\mathcal{A}+i\mathcal{C})-(\tilde{\mathcal{A}}+i\tilde{\mathcal{C}})|=\mathcal{O}(\sqrt{|\epsilon_1|+\epsilon_2}),
	\end{align}
	and are computable in classical $\mathcal{O}(\mathrm{poly(Q)})$ time.
\end{lemma}
\begin{proof}
	First, we satisfy condition (4) of ~\cref{Lem:AchievableAC} by rescaling 
	$
	\mathcal{A}_1=\frac{\tilde{ \mathcal{A}}}{1+\epsilon_2},
	\quad
	\mathcal{C}=\frac{\tilde{ \mathcal{C}}}{1+\epsilon_2}.
	$
	
	Second, we use the polynomial sum-of-squares technique to compute real Fourier series $\mathcal{B}_1=\sum_{k=0}^{Q/2}b_k\cos{(k\theta)}$ and $\mathcal{D}_1=\sum_{k=1}^{Q/2}d_k\sin{(k\theta)}$ such that $\mathcal{A}_1^2+\mathcal{B}_1^2+\mathcal{C}^2+\mathcal{D}_1^2=1$. Following~\cite[Section C]{Low2016}, this can be done in classical $\mathcal{O}(\mathrm{poly(Q)})$ time. By construction, $\mathcal{A}_1^2(0)+\mathcal{B}_1^2(0)=1$ as $\mathcal{C}$ and $\mathcal{D}_1$ are odd functions. Let us define $\sin{(\delta)}\equiv\mathcal{B}_1(0)$. Hence, $\cos{(\delta)}\equiv\mathcal{A}_1(0)=\frac{1+\epsilon_1}{1+\epsilon_2}$.
	
	Third, let $\mathcal{A}(\theta)=\cos{(\delta)}\mathcal{A}_1(\theta)+\sin{(\delta)}\mathcal{B}_1(\theta)$. By definition, $\mathcal{A}(0)=\cos^2{(\delta)}+\sin^2{(\delta)}=1$. Thus $\mathcal A,\mathcal C$ satisfy satisfy the conditions of~\cref{Lem:AchievableAC}.
	
	Last, we bound the error of this approximation.
	\begin{align}
	\nonumber&\max_{\theta\in\mathbb{R}}|(\mathcal{A}+i\mathcal{C})-(\tilde{\mathcal{A}}+i\tilde{\mathcal{C}})|\
	\\&
	\nonumber\le\max_{\theta\in\mathbb{R}}|(\mathcal{A}+i\mathcal{C})-(\mathcal{A}_1+i\mathcal{C})|+\max_{\theta\in\mathbb{R}}|(\mathcal{A}_1+i\mathcal{C})-(\tilde{\mathcal{A}}+i\tilde{\mathcal{C}})|
	& \text{by a triangle inequality}
	\\&
	\nonumber\le\max_{\theta\in\mathbb{R}}|(\cos{(\delta)}-1)\mathcal{A}_1(\theta)+\sin{(\delta)}\mathcal{B}_1(\theta)|+\epsilon_2
	\\&
	\nonumber\le\sqrt{(\cos{(\delta)}-1)^2+\sin^2{(\delta)}}+\epsilon_2	& \text{using}\;\mathcal{A}_1^2(\theta)+\mathcal{B}_1^2(\theta)\le 1
	\\&
	\nonumber\le \sqrt{2}\sqrt{1-\cos{(\delta)}}+\epsilon_2= \sqrt{2\frac{\epsilon_2-\epsilon_1}{1+\epsilon_2}}+\epsilon_2
	\\&
	=\mathcal{O}(\sqrt{|\epsilon_1|+\epsilon_2}).
	\end{align}
\end{proof}
We now prove~\cref{thm:AchievableAC} by evaluating~\cref{Eq.QSPcomponents} using~\cref{Lem:AchievableAC} for the case where the arbitrary unitary $\hat{V}$ is replaced with the more structured iterate $e^{i\Phi}\hat{W}$, and also where we have added a global phase $\Phi\in\mathbb{R}$.
\begin{proof}[Proof of~\cref{thm:AchievableAC}]
Observe from~\cref{Eq:W_Block_form} that $e^{i\Phi}\hat{W}$ can be diagonalized to obtain its eigenvalues $\theta_{\lambda\pm}$ and eigenvectors $\ket{G_{\lambda\pm}}$.
\begin{align}
\label{Eq:Iterate_Diagonalized}
e^{i\Phi}\hat{W}\ket{G_{\lambda\pm}}
=e^{i(\Phi+\theta_{\lambda\pm})}\ket{G_{\lambda\pm}}
,\quad
\theta_{\lambda\pm}=\mp \cos^{-1}{(\lambda)}
,\quad
\ket{G_{\lambda\pm}}=\frac{\ket{G_{\lambda}}\pm i\ket{G^{\perp}_{\lambda}}}{\sqrt{2}}.
\end{align} 
With this substitution $\hat{V}\leftarrow e^{i\Phi}\hat{W}$,~\cref{Eq.QSPcomponents} becomes
\begin{align}
\label{eq:Eq:QSP_iterate}
\hat{V}_{\vec{\varphi}}=\bigoplus_{\lambda,\pm}\left(
\I_b \mathcal A(\Phi+\theta_{\lambda\pm})
+i\hat{X}_b \mathcal  C(\Phi+\theta_{\lambda\pm})+\cdots \right)\otimes\ket{G_{\lambda\pm}}\bra{G_{\lambda\pm}}_{as}.
\end{align}
Similar to~\cref{Sec:Observable_Transformation}, we are only allowed to prepare and measure in the subspace supported by signal state $\ket{G}_a$. Recall that $(\bra{G}_a\otimes\I_s)\ket{G_\lambda^\perp}=0$, so projecting the sequence~\cref{eq:Eq:QSP_iterate} onto $\ket{+}_b\ket{G}_a$ results in
\begin{align}
\label{Eq:1QSPFullExpression}
\nonumber\bra{G}_a\bra{+}_b\hat{V}_{\vec\varphi}\ket{+}_b\ket{G}_a
&=
\bigoplus_{\lambda,\pm}\left(\mathcal A(\Phi+\theta_{\lambda\pm})
+i\mathcal  C(\Phi+\theta_{\lambda\pm}) \right) \otimes\bra{G}_a\ket{G_{\lambda\pm}}\bra{G_{\lambda\pm}}_{as}\ket{G}_a
\\
&=
\bigoplus_{\lambda}\frac{\sum_{\pm}\left(\mathcal A(\Phi+\theta_{\lambda\pm})
+i\mathcal  C(\Phi+\theta_{\lambda\pm}) \right)}{2} \otimes
\ket{\lambda}\bra{\lambda}.
\end{align}

For this proof, it suffices to choose $\Phi=\pi/2$, and set $a_k=0$ for all odd $k$, and $c_k=0$ for all even $k$. We then evaluate each component of~\cref{Eq:1QSPFullExpression} using~\cref{Lem:AchievableAC} and the Chebyshev polynomials $T_k(x)\equiv \cos{(k\cos^{-1}{(x)})}$. 
\begin{align}
&\mathcal{A}(\pi/2+\theta_{\lambda\pm})
=\sum_{k\;\text{even}}^{Q/2}a_k\cos{(k(\pi/2+\theta_{\lambda\pm}))}
=\sum_{k\;\text{even}}^{Q/2}a_k i^kT_k(\lambda)
=\sum_{k\;\text{even}}^{Q/2}a'_kT_k(\lambda)
=A(\lambda),
\\
\nonumber&\mathcal{C}(\pi/2+\theta_{\lambda\pm})
=\sum_{k\;\text{odd}}^{Q/2}c_k\sin{(k(\pi/2+\theta_{\lambda\pm}))}
=\sum_{k\;\text{odd}}^{Q/2}c_ki^{k-1}T_k(\lambda)
=\sum_{k\;\text{odd}}^{Q/2}c'_kT_k(\lambda)
=C(\lambda).
\end{align}
Thus~\cref{Eq:1QSPFullExpression} simplifies to 
\begin{align}
\label{Eq:1QSPSimpleExpression}
\bra{G}_a\bra{+}_b\hat{V}_{\vec\varphi}\ket{+}_b\ket{G}_a
=
\bigoplus_{\lambda}(A(\lambda)+iC(\lambda))\otimes
\ket{\lambda}\bra{\lambda}.
\end{align}

We now express the conditions of~\cref{Lem:AchievableAC} in terms of polynomials. As $T_k(-x)=(-1)^kT_k(x)$, conditions (1) and (2) map to $A(\lambda)$ being an even polynomial and $C(\lambda)$ being an odd polynomial respectively. When $\theta_{\lambda\pm}=-\pi/2$, this implies that $\lambda=\cos{(\pm\pi/2)}=0$. Hence condition (3) maps to $
\mathcal{A}(\pi/2+\theta_{\lambda\pm}=0)=1 \Rightarrow A(\lambda =0)=1.
$
As $\lambda\in[-1,1]$ for all values of $\theta_{\lambda\pm}$, condition (4) maps to
$\forall \lambda \in[-1,1], A^2(\lambda)+C^2(\lambda)\le 1$.
\end{proof}

Of course, other choices of $\Phi$, such as $\Phi=0,\pi/3,\pi/4,...$, lead to different families of target functions. However, those are beyond the scope of this work.

\section{Application to Hamiltonian Simulation}
\label{Sec:Hamiltonian_Simulation}
With these results for qubitization and operator function design with a quantum signal processor, the application to Hamiltonian simulation follows easily. We complete the proof~\cref{thm:QueryComplexity}, and then apply these results to obtain our claims of improvements in~\cref{Tab:Comparison}. Given any Hamiltonian $\hat{H}$ encoded in the standard-form of~\cref{def:standardform},~\cref{thm:AchievableAC} allows us to use $2Q$ queries to encode exactly any function $A[\hat{H}]+iC[\hat{H}]$ where $A(\lambda)$ and $C(\lambda)$ are bounded polynomials of opposite parity and degree $Q$. Hamiltonian simulation is accomplished by choosing a good degree $Q$ polynomial approximation to $A(\lambda)+iC(\lambda)=e^{-i\lambda t}$.
\begin{proof}[Proof of~\cref{thm:QueryComplexity}]
Similar to previous approaches~\cite{Berry2015Hamiltonian,Low2016HamSim}, we approximate $e^{-i\lambda t}$ with the Jacobi-Anger expansion $e^{i\cos{(z)}t}=\sum^{\infty}_{k=-\infty}i^k J_k(t)e^{i k z}$~\cite{Abramowitz1966handbook}, where $J_k(t)$ are Bessel function of the first kind. By identifying $\cos{(z)}=\lambda$, suitable polynomials $A,C$ are obtained by a truncation and rescaling of
\begin{align}
e^{-i\lambda t}=J_0(t)+2\sum^\infty_{k\;\text{even}> 0}(-1)^{k/2}J_{k}(t)T_{k}(\lambda)+i2\sum^\infty_{k\;\text{odd}>0}(-1)^{(k-1)/2}J_{k}(t)T_{k}(\lambda).
\end{align}
The error from truncating this expansion for $k > Q=q-1$ is a sum of $|J_k(t)|$ that was bounded in~\cite{Berry2015Hamiltonian}:
\begin{align}
\label{eq:TruncationError}
&\epsilon
=
\max_{\lambda\in[-1,1]} \left|2\sum^\infty_{k=q}(-1)^{\frac{k}{2}}J_{k}(t)T_{k}(\lambda)\right|
\le 
\sum^{\infty}_{k=q}\displaystyle 2|J_{k}(t)|\le \frac{4t^{q}}{2^{q} q!}=\mathcal{O}\Big(\Big(\frac{e t}{q}\Big)^{q}\Big)
\\
\nonumber
&
\Rightarrow
\log{\left(\frac{1}{\epsilon}\right)}=\mathcal{O}\left(q\log{\left(\frac{2q}{e t}\right)}\right)
\\\nonumber
&\Rightarrow q=\mathcal{O}(t+\log{(1/\epsilon)}).
\end{align}
The rapid converge by truncation arises as $e^{-i\lambda t}$ is an entire analytic function~\cite{Boyd2007Sine}. 

With the choice
\begin{align}
\label{eq:jacobi-anger-truncation}
\tilde{A}(\lambda)=J_0(t)+2\sum^Q_{k\;\text{even}> 0}(-1)^{k/2}J_{k}(t)T_{k}(\lambda)
,\quad
\tilde{C}(\lambda)=2\sum^Q_{k\;\text{odd}>0}(-1)^{(k-1)/2}J_{k}(t)T_{k}(\lambda)
\end{align}
the error $\max_{\lambda\in[-1,1]}|\tilde A(\lambda)+i\tilde C(\lambda)- e^{-i \lambda t}|\le\epsilon$. Though this choice satisfies conditions (1-2) of~\cref{thm:AchievableAC}, it is only $\epsilon$-close to satisfying conditions (3-4). Fortunately, this is not a fundamental problem following the stability analysis of~\cref{Lem:StabilityAC}. One can perturb $\tilde{A}(\lambda), \tilde{C}(\lambda)$ to construct approximations $A(\lambda)+iC(\lambda)$ that are achievable, with error $\max_{\lambda\in[-1,1]}|A(\lambda)+iC(\lambda)- e^{-i \lambda t}|=\mathcal{O}(\sqrt{\epsilon)}$. Thus the overall procedure constructs a unitary $\hat{V}_{\vec\varphi}$ that approximates the time-evolution operator with error $\|\bra{+}_b\bra{G}_a\hat{V}_{\vec\varphi}\ket{G}_a\ket{+}_b-e^{-i\hat{H}t}\| = \mathcal{O}(\sqrt{\epsilon})$. The failure probability is $\mathcal{O}(\epsilon)$, and solving for $Q$ furnishes stated the number of queries to $\hat{W}$ and hence the oracles encoding $\hat{H}=\bra{G}\hat{U}\ket{G}$.
\end{proof}

As we have already shown in~\cref{sec:standardform} how a variety of matrix input models such as linear-combination-of-unitaries, $d$-sparse Hamiltonians, and purified density matrices map to the standard-form encoding, their respective algorithms for simulating the time-evolution operators follow trivially.
\begin{corollary}[Hamiltonian Simulation of a Sparse Hermitian Matrix]
	\label{cor:HamSparse}
	Given access to the oracle of~\cref{eq:oracles_sparse} specifying a $d$-sparse Hamiltonian $\hat{H}$, time evolution by $\hat{H}$ can be simulated for time $t$ and error $\epsilon$ with $\mathcal{O}(d t\|\hat{H}\|_{\text{max}}+\log{(1/\epsilon))}$ queries to $\hat{O}_H,\hat{O}_F$.
\end{corollary}
\begin{proof}
	Follows from combining~\cref{thm:QueryComplexity} with the standard-form encoding~\cref{cor:encoding_sparse}. 
\end{proof}
The query complexity in~\cref{eq:TruncationError} for this sparse case exactly matches a lower bound based on simulating a Hamiltonian that solves PARITY with unbounded error~\cite{Berry2014Exponential}, valid for all parameter values, and not just in asymptotic limits~\cite{Berry2015Hamiltonian,Low2016HamSim}. This completes the optimality proof of~\cref{thm:QueryComplexity}. 

The case where $\hat{H}$ decomposes into a linear combination of unitaries is an immediate application:
\begin{corollary}[Hamiltonian Simulation of a Linear Combination of Unitaries]
	\label{cor:HamSum}
	Given access to the oracles of~\cref{Eq:AV_definition} specifying a Hamiltonian $\hat{H}=\sum^d_{j=1}\alpha_j\hat{U}_j$, time evolution by $\hat{H}$ can be simulated for time $t$ and error $\epsilon$ with $\mathcal{O}(\alpha t+{\log{(1/\epsilon)}})$ queries.
\end{corollary}
\begin{proof}
	Follows from combining~\cref{thm:QueryComplexity} with the standard-form encoding~\cref{cor:encoding_LCU}. 
\end{proof}

The intuitiveness of~\cref{thm:QueryComplexity} allows us to swiftly devise new models of Hamiltonian simulation.
\begin{corollary}[Hamiltonian Simulation of a Purified Density Matrix]
	\label{cor:HamPure}
	Given access to the oracle~\cref{eq:oracle_density_matrix} specifying a Hamiltonian $\hat{H}=\hat{\rho}$ that is a density matrix $\hat{\rho}$, time evolution by $\hat{H}$ can be simulated for time $t$ and error $\epsilon$ with $\mathcal{O}(t+\log{(1/\epsilon)})$ queries.
\end{corollary}
\begin{proof}
	Follows from combining~\cref{thm:QueryComplexity} with the standard-form encoding~\cref{cor:encoding_purified}. 
\end{proof}

In all the above, the query complexity $\mathcal{O}(\cdots t+\log{(1/\epsilon)})$ is an upper bound on the more precise form of~\cref{eq:TruncationError}. The exact the tradeoff between $\epsilon,t,Q$ in~\cref{eq:TruncationError} is studied numerically in Appendix~\ref{Sec:ApplicationOfHamSim}, together with example phases $\vec{\varphi}$ implementing $\hat{V}_{\vec\varphi}$ for the polynomials in~\cref{eq:jacobi-anger-truncation}. 

\section{Conclusion}
Our general procedure for Hamiltonian simulation in~\cref{thm:QueryComplexity}  extends the scope of possible useful formulations of Hamiltonian simulation. As seen in~\cref{Tab:Comparison}, it encompasses any case where the Hamiltonian is embedded in a flagged subspace of the signal unitary. Given this, a simulation algorithm with query complexity optimal in all parameters, and also not just in asymptotic limits, is easily obtained with minimal overhead. While this procedure contains and significantly improves upon important models where the Hamiltonian is $d$-sparse or a linear combination of unitaries, its greater value lies in illuminating an intuitive and straightforward path to other as-yet undiscovered models of Hamiltonian simulation. In particular, our result for time-evolution by a purified density matrix is a quadratic improvement in time and an exponential improvement in error over the sample-based model -- the proof of which consisted of just a few lines.

Many other exciting directions extend from this work. One example is how additional structural information about $\hat{H}$~\cite{Childs2010Limitation} may be exploited. This is illustrated by when the spectral norm of $\|\hat{H}\|$ is significantly smaller than the sum of coefficients $\|\vec{\alpha}\|_1$ of a particular linear combination of unitaries decomposition. If this decomposition were to be used, simulation would take time $\mathcal{O}\big( \|\vec{\alpha}\|_1t+\log{(1/\epsilon)}\big)$ -- a factor $\|\vec{\alpha}\|_1/ \|\hat{H}\|$ slowdown. Unfortunately, an $\|\vec{\alpha}\|_1=\mathcal{O}(\|\hat{H}\|)$ decomposition appears difficult to find in general. Furthermore, it is easy to construct pathological $\hat{H}$ with small norms, but nevertheless decompose by naive methods into components with large spectral norms~\cite{Somma2016Trotter}. Our approach offers a possible solution -- one finds any encoding the Hamiltonian in the standard-form, rather than some specific decomposition into parts with properties dictated by the formulation. It remains an interesting challenge to identify the cases where, and determine how structural information may be incorporated.

These advances are special cases arising from our vision of the more general quantum signal processor. Through qubitization, structure is imposed onto any unitary process implementing some Hermitian signal operator. This structure allows for efficient processing of the signal by the techniques of quantum signal processing into some more desired form. As the query complexity of approximating any arbitrary target function of the signal exactly matches fundamental bounds lower in polynomial and Fourier approximation theory~\cite{Meinardus1967}, we expect this to have numerous application in metrology~\cite{Low2015} and other quantum algorithms~\cite{Yoder2014}.

\subsection{Developments after preprint release}
The preprint of this manuscript was released on October 2016. There have since been a number of developments building off, or relating to, the concepts of the standard-form encoding, qubitization, and quantum signal processing that could be of interest to the reader.
\begin{itemize}
	\item \textbf{Singular value transformations}: 
	The standard-form encoding, qubitization, and quantum signal processing of Hermitian or normal matrices has been extended to perform polynomial transformations on the singular value of arbitrary matrices~\cite{Gilyen2018singular}. This generalization has proven to be surprisingly broad, and provides a unified framework for deriving and understanding many other quantum algorithms.
	\item \textbf{Structured Hamiltonian simulation}: The Hamiltonian of many interesting systems possess additional structure that should be exploited to further reduce the cost of simulation. More refined simulation algorithms can exploit prior knowledge of geometric locality~\cite{Haah2018quantum, Childs2019Lattice}, Hamiltonian induced one-norm~\cite{Low2017USA}, spectral norm~\cite{Low2018SpectralNorm}, and separation of energy scales~\cite{Low2018IntPicSim}. In particular, these algorithms are also optimal, up to sub-polynomial factors, with respect to their structural parameters.
	%\item \textbf{Case studies and application}: A number of case studies and applications incorporating various ideas in this work can be found. This includes the simulation of Heisenberg chains~\cite{Childs2017Speedup}, chemistry~\cite{Babbush2018encoding, Low2019qSharp}, algorithms for linear systems of equations~\cite{Chakraborty2018BlockEncoding}, and spectral measurements~\cite{Poulin2018Spectral,babbush2017sorting}.
	\item \textbf{Numerical Quantum signal processing}: The numerical algorithm for decomposing target polynomials of degree $Q$ into phases was originally claimed to be polynomial time~\cite{Low2016}, but only when counting arbitrary-precision arithmetical operations, and appeared to be ill-conditioned in case studies~\cite{Childs2017Speedup}. A more thorough analysis~\cite{Haah2018product}  proved that the algorithm had a runtime of $\mathcal{O}(Q^3 \operatorname{polylog}(Q/\epsilon))$ with finite-precision arithmetic, provided that some subtleties were carefully managed to control this ill-conditioned behavior.
\end{itemize}

\section{Acknowledgments}
G.H.Low and I.L.Chuang thank Robin Kothari, Yuan Su, and Andrew Childs for insightful discussions, the excellent reviews by anonymous referees, and acknowledge funding by the ARO Quantum Algorithms Program and NSF RQCC Project No.1111337.
\appendix
\section{Qubitization of Normal Operators}
\label{Sec:NormalOperators}

The results of~\cref{Thm.Existence_of_S,Thm:Subspace} for qubitization can be extended to normal operators. Let $\bra{G}_a\hat{U}\ket{G}_a=\hat{H}$ in~\cref{Eq:HermitianTransformation} encode a normal matrix $\hat{H}$. It is well known that any normal matrix has a polar decomposition
\begin{align}
\hat{H}=\hat{H}_U \cdot \hat{H}_H.
\end{align}
where $\hat{H}_U$ is unitary, $\hat{H}_H$ is positive-semidefinite, $[\hat{H}_U,\hat{H}_H]=0$ commute. Thus the eigenvalues are $\hat{H}\ket{\lambda}=e^{i\theta_\lambda}\lambda\ket{\lambda}$, where $\lambda\ge 0, \;\theta_{\lambda}\in\mathbb{R}$, and 
\begin{align}
\label{Eq:NormalTransformationEigenstate}
\hat{U}\ket{G}\ket{\lambda}&=\hat{U}\ket{G_\lambda}=\lambda e^{i\theta_\lambda} \ket{G_\lambda}+\sqrt{1-|\lambda|^2}\ket{G_\lambda^{\perp}},
\end{align}

This reduces to a Hermitian operator when $\hat{H}_U$ has eigenvalues $\pm 1$, and reduces to a unitary operator when all $\hat{H}_H$ has eigenvalues $1$. The trivial approach to qubitization in fact applies to any complex matrix. We simply use the construction of~\cref{Thm.Existence_of_S} to implement the Hermitian signal operator $\frac{1}{2}(\hat{H}+\hat{H}^{\dag})$.

Another possibility uses two phased iterates in an alternating sequence on input state $\ket{G}_a\ket{\psi}_s$,
\begin{align}
\hat{W}_{\phi\pm}&=\hat{Z}_{\phi-\pi/2}(2\ket{G}\bra{G}-\I)\hat{U}_\pm\hat{Z}_{-\phi+\pi/2},
\quad
\hat{W}_{\vec{\phi}\pm}=\hat{W}_{\vec{\phi}}=\hat{W}_{\phi_L+}\cdots \hat{W}_{\phi_3+}\hat{W}_{\phi_2-}\hat{W}_{\phi_1+},
\end{align}
where $\hat{U}_+ = \hat{U}$ and $\hat{U}_- = \hat{U}^{\dag}$. We work through a minimal example in~\cref{sec:normal_example}, from which it follows that for each eigenstate $\ket{\lambda}$, the iterates
$\hat{W}_{\phi\pm}$ maps states in the subspace spanned by $\{\ket{G}\ket{\lambda},\hat{W}_{\pi/2,\pm}\ket{G}\ket{\lambda}\}$ to a state in the span of $\{\ket{G}\ket{\lambda},\hat{W}_{\pi/2,\mp}\ket{G}\ket{\lambda}\}$. Let us define the following states by Gram-Schmidt orthogonalization
\begin{align}
\ket{G_\lambda^{\perp \pm}}=\frac{(\lambda e^{i\theta_\lambda}-\hat{W}_{\pi/2,\pm})\ket{G_\lambda}}{\sqrt{1-|\lambda|^2}}.
\end{align}
Then these iterates have form
\begin{align}
\hat{W}_{\phi\pm}=
\begin{array}{rr}
	\lambda e^{\pm i\theta_\lambda}\ket{G_\lambda}\bra{G_\lambda}
	&-ie^{-i\phi}\sqrt{1-\lambda^2}\ket{G_\lambda}\bra{G^{\perp\mp}_\lambda}\\
	-ie^{i\phi}\sqrt{1-\lambda^2}\ket{G^{\perp\pm}_\lambda}\bra{G_\lambda}
	&+\lambda e^{\mp i\theta_\lambda}\ket{G^{\perp\pm}_\lambda}\bra{G^{\perp\mp}_\lambda}
\end{array}.
\end{align}
The subspace spanned by these states is only invariant in general under the action of iterates with alternating sign $\pm$, such as the product  $\hat{W}_{\phi_{i}-}\hat{W}_{\phi_j+}$. In other words, they are not invariant under any repeated application of an iterate of the same sign.
With the understanding that we only consider alternating sequences, each $\hat{W}_{\phi\pm}$ has the representation 
\begin{align}
\hat{W}_{\phi\pm}=
\left(\begin{matrix}
e^{\pm i\theta_\lambda}\lambda & -ie^{-i\phi}\sqrt{1-\lambda^2} \\
-ie^{i\phi}\sqrt{1-\lambda^2} & e^{\mp i\theta_\lambda}\lambda
\end{matrix}\right).
\end{align}
Note that when all eigenvalues $\lambda$ are identical and $\phi=\pi/2$, this reduces to Oblivious amplitude amplification~\cite{Berry2012}, and we recover Hermitian qubitization when all $\theta_\lambda=0$. While this approach uses one less ancilla qubit than the construction of~\cref{Thm.Existence_of_S}, quantum signal processing can only be performed on controlled block of even length $\hat{W}_{\vec{\phi}\pm}$, as only they have an invariant subspace. This limitation can be relevant in some cases, such as Hamiltonian simulation where quantum signal processing is applied to a single $\hat{W}_\phi$.

\subsection{Minimal example}
\label{sec:normal_example}
A minimal example illustrating the phenomenon of normal qubitization is worked through in this section.
For simplicity, let us drop the $\ket{\lambda}$ state. Let us set the phase $\phi=\pi/2$ in the $\hat{Z}_\cdot$ operator. Let $\ket{G}\rightarrow\ket{0}$. Let $R=2\ket{0}\bra{0}-1$. Let us map the subspace spanned by $\{\ket{G}\ket{\lambda},\ket{G_\lambda^{\perp +}},\ket{G_\lambda^{\perp -}}\}\rightarrow \{\ket{0},-\ket{1},-\ket{2}\}$. Then
\begin{align}
U\ket{0}&=\lambda\ket{0}+\sqrt{1-|\lambda|^2}\ket{1}, \quad U^\dag \ket{0}=\lambda^*\ket{0}+\sqrt{1-|\lambda|^2}\ket{2},
\\\nonumber
RU\ket{0}&=\lambda\ket{0}-\sqrt{1-|\lambda|^2}\ket{1}, \quad RU^\dag \ket{0}=\lambda^*\ket{0}-\sqrt{1-|\lambda|^2}\ket{2}.
\end{align}
Note that $\lambda\in\mathbb{C}$. Note that $\stab{1|2}\neq 0$ in general. Also note that $\bra{0}URU^\dag\ket{0}=\bra{0}U^\dag RU\ket{0}=2|\lambda|^2-1$ by an explicit calculation.

By Gram-Schmidt orthoormalization,  
\begin{align}
\ket{1}=\frac{(U-\lambda)\ket{0}}{\sqrt{1-|\lambda|^2}},\quad \ket{2}=\frac{(U^\dag-\lambda^*)\ket{0}}{\sqrt{1-|\lambda|^2}}
\end{align}
Promised that the first input state is $\ket{0}$, and we only consider sequences of the form $(RU^\dag RU)^k$ and $RU(RU^\dag RU)^k$ for integer $k$, we want to prove that any subsequent input to $RU$ is in $\s\{\ket{0},\ket{2}\}$, and any input to $RU^\dag$ is in $\s\{\ket{0},\ket{1}\}$. Consider the projectors
%\begin{align}
$P_j=\ket{0}\bra{0}+\ket{j}\bra{j}$.
%\end{align}
Then proving the first case requires showing that $\bra{1}UR P_2 RU^\dag\ket{1}=1$ and for the second case, $\bra{2}U^\dag R P_1 RU\ket{2}=1$. In the first case,
\begin{align}
\bra{1}UR P_2 RU^\dag\ket{1}&=\frac{|\bra{0}RU^\dag(U-\lambda)\ket{0}|^2}{1-|\lambda|^2}+\frac{|\bra{0}(U-\lambda)RU^\dag(U-\lambda)\ket{0}|^2}{(1-|\lambda|^2)^2}
\\\nonumber
&=
\frac{|1-|\lambda|^2|^2}{1-|\lambda|^2}+\frac{|\lambda|\lambda|^2-\lambda(2|\lambda|^2-1)-\lambda+\lambda|^2}{(1-|\lambda|^2)^2}
\\\nonumber
&=
\frac{|1-\lambda^2|^2}{1-\lambda^2}+\frac{\lambda^2|1-\lambda^2|^2}{(1-\lambda^2)^2}=1.
\end{align}
By an identical calculation, the second case is also true.

This means that $RU$ maps any state in $\s\{\ket{0},\ket{2}\}$ to a state in $\s\{\ket{0},\ket{1}\}$, and $RU^\dag$ maps any state in $\s\{\ket{0},\ket{1}\}$ to a state in $\s\{\ket{0},\ket{2}\}$. Using the fact that $\bra{0}RU\ket{2}=\bra{2}U^\dag\ket{0}^*=\sqrt{1-|\lambda|^2}$, we know that the overlap $|\bra{1}RU\ket{2}|=|\lambda|$. With a more careful evaluation,
\begin{align}
\bra{1}RU\ket{2}=\frac{\bra{0}(U^\dag-\lambda^*)RU(U^\dag-\lambda^*)\ket{0}}{1-|\lambda|^2}=\frac{\lambda^*|\lambda|^2-\lambda^*(2|\lambda|^2-1)-\lambda^*+\lambda^*}{1-|\lambda|^2}=\lambda^*.
\end{align} 
This means that we can represent
\begin{align}
RU&=(\lambda\ket{0}-\sqrt{1-|\lambda|^2}\ket{1})\bra{0}+(\sqrt{1-|\lambda|^2}\ket{0}+\lambda^*\ket{1})\bra{2}
\\\nonumber
&
=\begin{array}{rr}
\lambda\ket{0}\bra{0} &   +\sqrt{1-|\lambda|^2}\ket{0}\bra{2} \\
-\sqrt{1-|\lambda|^2}\ket{1}\bra{0}& +\lambda^*\ket{1}\bra{2}
\end{array},
\end{align}
and similarly for $RU^\dag$.
\begin{figure}[t]
	\centering
	\parbox{0.48\linewidth}{
		\includegraphics[width=\linewidth]{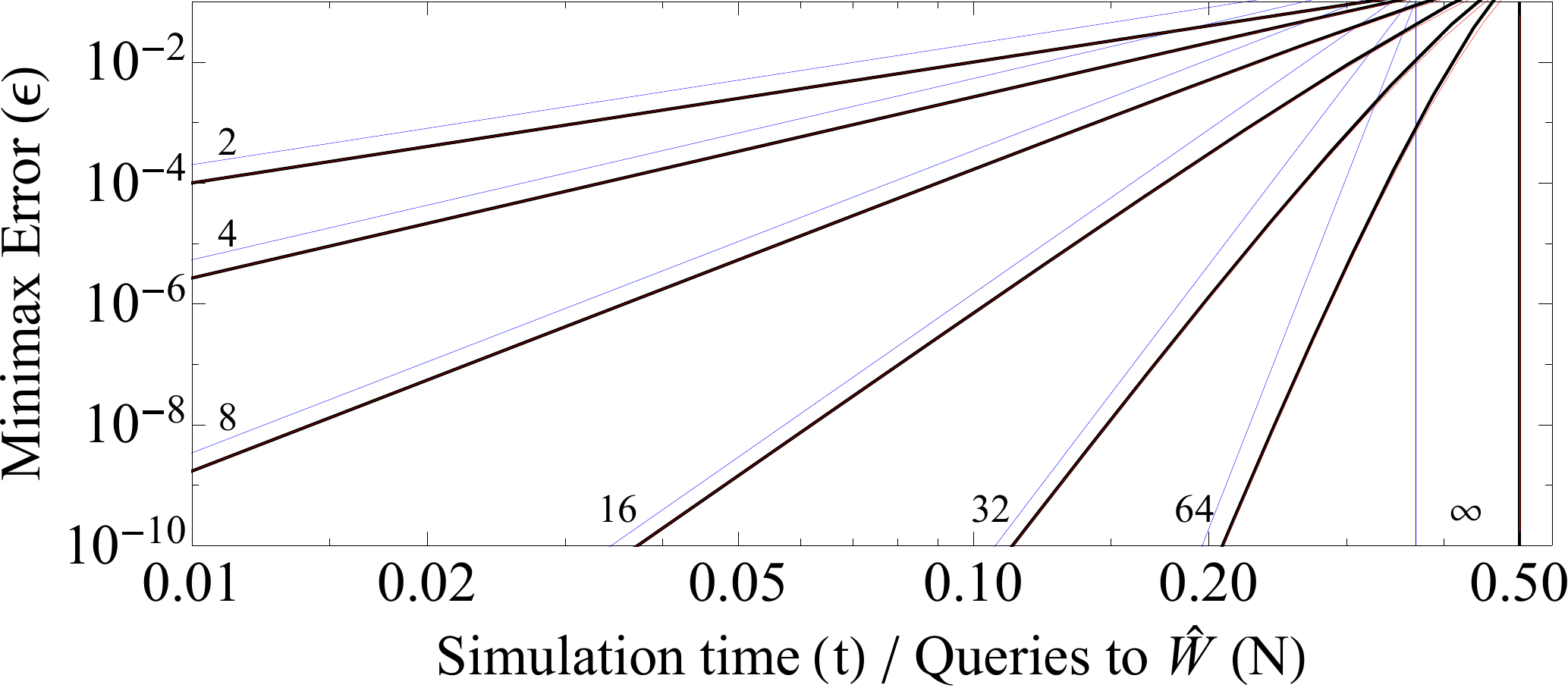}
	}
	\quad
	\parbox{0.48\linewidth}{
		\includegraphics[width=1.0\linewidth]{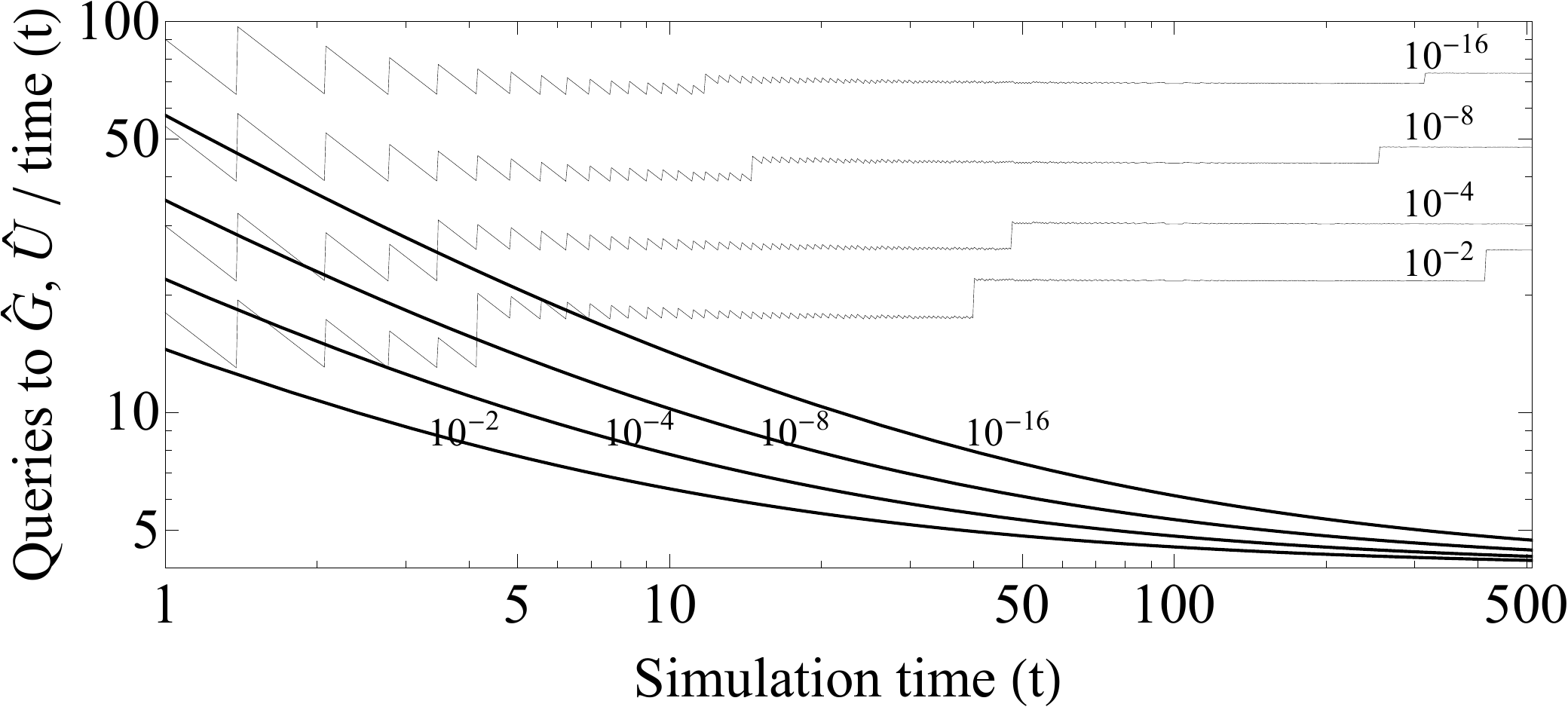}
	}
	\caption{
		\label{fig:AppendixScalingPlot}
		(Left) Approximation error $\epsilon=\max_{\theta\in\mathbb{R}}|A[\theta]-iC[\theta]-e^{i t \sin{(\theta)}}|$. $(A[\theta],C[\theta])$ are real Fourier series in $(\cos{(k\theta)},\sin{(k\theta)}),\;k=0,...,Q/2$, and $\epsilon$ is plotted for the upper bound $\frac{4t^q}{2^q q!}$ (blue), truncation $\sum^{\infty}_{k=q} 2|J_{k}[t]|$ (black), and best possible~\cite{karam1999} (red), for $Q=2,4,8,16,32,64,\infty$ queries to the controlled iterate $\hat{W}$, where $q=1+Q/2$. (Right) Queries per unit of simulation time to unitary $\hat{U}$ encoding $\hat{H}=\bra{G}\hat{U}\ket{G}$  at target error $\|\bra{+}\hat{V}_{\vec{\varphi}}\ket{+}-e^{-i\hat{H}t}\|=10^{-2,-4,-8,-16}$ for the BCCKS algorithm (thin), and this work using the truncated approximation to $e^{i t \sin{(\theta)}}$ (thick).
	}
\end{figure}

\section{Practical Details for Implementing Hamiltonian Simulation}
\label{Sec:ApplicationOfHamSim}
This appendix illustrates a specific application of the quantum signal processing approach to a signal unitary that encodes the Hamiltonian $\hat{H}$ as a signal operator. In particular, a comparison of performance with the BCCKS approach is made. The details will be useful to readers interested in implementing our procedure on a quantum computer. These include plots for the exact error scaling~\cref{eq:TruncationError} of the $Q/2$ term Fourier approximation to $e^{i t\sin{(\theta)}}$ in~\cref{fig:AppendixScalingPlot}(Left), and the number of required queries per unit of simulated time in~\cref{fig:AppendixScalingPlot}(Right). Furthermore, a table of select phases $\vec{\varphi}$ computed using the algorithm in~\cite{Low2016} can be found in~\cref{Tab:Phases}. %For completeness, we repeat some details 
\begin{table}[t]
	\begin{tabularx}{\linewidth}{c|c|c|X}
		$Q$ & $\epsilon$ & $t$ &  $\vec{\varphi}=(\varphi_1,\varphi_2,...,\varphi_Q)$ for $h(\theta)=t\sin{(\theta)}$\\
		\hline
		\hline
		2 & $10^{-2}$ & 0.0707 & (-1.61, 1.67) \\ \hline
		4 & $10^{-2}$ & 0.311 & (-1.03, 2.54, 1.36, -2.11) \\ \hline
		8 & $10^{-2}$ & 1.20 & (-2.63, 1.66, -2.42, -1.79, 2.17, 1.43, -2.71), \\ \hline
		16 & $10^{-2}$ & 3.78 & (0.23, 1.17, 3.07, -1.63, -1.78, 2.88, 1.71, 2.77, -1.95, 3.12, 1.97, -2.56, -2.87, 2.00, -2.29, 2.92, -0.51) \\ \hline
		32 & $10^{-2}$ & 10.1 & (-2.94, 2.64, -2.32, 2.42, -2.86, -2.72, 2.4, -2.57, -2.96, 2.43, -2.53, -2.63, 2.33, 3.11, -2.17, 3.07, 2.24, -2.98, -1.99, -2.77, 2.22, 2.33, -2.62, -1.75, -2.15, 2.84, 1.68, \
		1.48, 2.46, -2.26, -0.92, -0.21) \\ \hline
		2 & $10^{-4}$ & 0.0070711 & (-1.574, 1.5817) \\ \hline
		4 & $10^{-4}$ & 0.066948 & (-0.6741, 2.5806, 0.7772, -2.4675) \\ \hline
		8 & $10^{-4}$ & 0.47498 & (-0.0914, -1.2861, 2.1995, 0.995, -2.637, -1.5369, 1.9236, -3.0502) \\ \hline
		16 & $10^{-4}$ & 2.2164 & (-0.5228, -2.9239, 1.3204, 2.6065, -1.73, -2.6191, 1.7225, 3.0121, -1.2286, -2.5276, 1.8058, -0.4605, 1.1563, -2.3492, 2.1623, -2.6188) \\ \hline
		32 & $10^{-4}$ & 7.3957 & (1.3594, -2.7039, -1.5041, -1.0845, 2.4817, 2.5571, -3.0846, 1.0661, 2.6256, -2.0267, -2.0383, 3.0857, 1.9338, 2.6966, -2.1759, -2.562, 2.2839, 2.6493, -2.2281, -2.9435, 2.1726, -2.9934, -2.9327, 1.7167, -3.0853, -0.9383, -0.2802, -0.2376, -2.9556, 2.7875, -2.2875, 1.7822) \\\hline
	\end{tabularx}
	\caption{\label{Tab:Phases}Table of phases implementing target function $h(\theta)=t\sin{(\theta)}$ in quantum signal processing. Errors quoted refer to $\|\bra{+}_b\hat{V}_{\vec{\varphi}}\ket{+}_b-e^{-i\hat{H}t}\|\le \epsilon$.}
\end{table}

In the interests of a fair comparison,~\cref{fig:AppendixScalingPlot}(Right) counts the number of queries to the signal operator $\hat{U}$ in the $\hat{H}=\bra{G}\hat{U}\ket{G}$ encoding. $\hat{U}$ is not assumed to be qubitized, thus incurring a factor $2$ additional cost from querying  $\hat{U}$ and $\hat{U}^{\dag}$ each over the asymptotic limit of $2$ queries per unit of simulation time in~\cref{fig:AppendixScalingPlot}(Left). Similarly, the BCCKS algorithm~\cite{Berry2015Truncated} incurs a factor $3$ additional cost from querying $\hat{U}$ twice and $\hat{U}^{\dag}$ in their use of oblivious amplitude amplification. As BCCKS is known to be optimal in the regime $t=\mathcal{O}(\frac{\log{(1/\epsilon)}}{\log\log{(1/\epsilon)}})$, the improvement of our approach is most dramatic outside of it. In particular, the queries per unit time of BCCKS scales like $\mathcal{O}(\frac{\log{(t/\epsilon)}}{\log\log{(t/\epsilon)}})$, whereas our approach approaches $4$ in the limit $t\rightarrow \infty$. 
%Another approach based on correcthat imp BCCKS refinement of corrected quantum walks was proposed ~\cite{Berry2016corrected}

The number of queries to in the BCCKS model is $3Kr$, where $K$ is the queries per segment $e^{-i\hat{H}\log{2}/r}$, $r=\lceil t/\log{(2)}\rceil$ is the number of segments, and $3$ is for oblivious amplitude amplification on each segment. $K$ is chosen such that $\sum^{\infty}_{k=K+1}\frac{\log^k{2}}{k!}\le\frac{\epsilon}{r}$. An analysis of the procedure~\cite{Berry2015Truncated} shows that the error of its simulated evolution is a factor $\mathcal{O}(1)\gtrsim 2$ larger than $\epsilon$. Thus~\cref{fig:AppendixScalingPlot} slightly overestimates BCCKS performance as we take $\epsilon$ directly to be the error.
%=3\argmin_{K}(\sum^{\infty}_{k=K+1}\frac{\log^k{2}}{k!})$

\bibliographystyle{apsrev4-1}
\nocite{apsrev41Control}
\bibliography{apsrev-control,QubitizationB}

\end{document}